\documentclass[aps,prd,twocolumn,superscriptaddress,showpacs,floatfix,letterpaper]{revtex4}
\usepackage{graphicx}
\usepackage{amsmath, amssymb}
\usepackage{dcolumn}
\usepackage{bm}
\usepackage{xspace}
\usepackage{multirow}
\usepackage{booktabs}
\usepackage{longtable}
\usepackage{textcomp}
\usepackage{subfigure}
\usepackage{enumerate}
\input{symblibrary.tex}
\begin{document} 
%
%
%
%
\title{ 
  Measurement of the masses and widths of the bottom baryons
  \( \mathbf{\Sigbpm} \) and \( \mathbf{\Sigbstpm} \) 
}
\affiliation{Institute of Physics, Academia Sinica, Taipei, Taiwan 11529, Republic of China}
\affiliation{Argonne National Laboratory, Argonne, Illinois 60439, USA}
\affiliation{University of Athens, 157 71 Athens, Greece}
\affiliation{Institut de Fisica d'Altes Energies, ICREA, Universitat Autonoma de Barcelona, E-08193, Bellaterra (Barcelona), Spain}
\affiliation{Baylor University, Waco, Texas 76798, USA}
\affiliation{Istituto Nazionale di Fisica Nucleare Bologna, $^{ee}$University of Bologna, I-40127 Bologna, Italy}
\affiliation{University of California, Davis, Davis, California 95616, USA}
\affiliation{University of California, Los Angeles, Los Angeles, California 90024, USA}
\affiliation{Instituto de Fisica de Cantabria, CSIC-University of Cantabria, 39005 Santander, Spain}
\affiliation{Carnegie Mellon University, Pittsburgh, Pennsylvania 15213, USA}
\affiliation{Enrico Fermi Institute, University of Chicago, Chicago, Illinois 60637, USA}
\affiliation{Comenius University, 842 48 Bratislava, Slovakia; Institute of Experimental Physics, 040 01 Kosice, Slovakia}
\affiliation{Joint Institute for Nuclear Research, RU-141980 Dubna, Russia}
\affiliation{Duke University, Durham, North Carolina 27708, USA}
\affiliation{Fermi National Accelerator Laboratory, Batavia, Illinois 60510, USA}
\affiliation{University of Florida, Gainesville, Florida 32611, USA}
\affiliation{Laboratori Nazionali di Frascati, Istituto Nazionale di Fisica Nucleare, I-00044 Frascati, Italy}
\affiliation{University of Geneva, CH-1211 Geneva 4, Switzerland}
\affiliation{Glasgow University, Glasgow G12 8QQ, United Kingdom}
\affiliation{Harvard University, Cambridge, Massachusetts 02138, USA}
\affiliation{Division of High Energy Physics, Department of Physics, University of Helsinki and Helsinki Institute of Physics, FIN-00014, Helsinki, Finland}
\affiliation{University of Illinois, Urbana, Illinois 61801, USA}
\affiliation{The Johns Hopkins University, Baltimore, Maryland 21218, USA}
\affiliation{Institut f\"{u}r Experimentelle Kernphysik, Karlsruhe Institute of Technology, D-76131 Karlsruhe, Germany}
\affiliation{Center for High Energy Physics: Kyungpook National University, Daegu 702-701, Korea; Seoul National University, Seoul 151-742, Korea; Sungkyunkwan University, Suwon 440-746, Korea; Korea Institute of Science and Technology Information, Daejeon 305-806, Korea; Chonnam National University, Gwangju 500-757, Korea; Chonbuk National University, Jeonju 561-756, Korea}
\affiliation{Ernest Orlando Lawrence Berkeley National Laboratory, Berkeley, California 94720, USA}
\affiliation{University of Liverpool, Liverpool L69 7ZE, United Kingdom}
\affiliation{University College London, London WC1E 6BT, United Kingdom}
\affiliation{Centro de Investigaciones Energeticas Medioambientales y Tecnologicas, E-28040 Madrid, Spain}
\affiliation{Massachusetts Institute of Technology, Cambridge, Massachusetts 02139, USA}
\affiliation{Institute of Particle Physics: McGill University, Montr\'{e}al, Qu\'{e}bec, Canada H3A~2T8; Simon Fraser University, Burnaby, British Columbia, Canada V5A~1S6; University of Toronto, Toronto, Ontario, Canada M5S~1A7; and TRIUMF, Vancouver, British Columbia, Canada V6T~2A3}
\affiliation{University of Michigan, Ann Arbor, Michigan 48109, USA}
\affiliation{Michigan State University, East Lansing, Michigan 48824, USA}
\affiliation{Institution for Theoretical and Experimental Physics, ITEP, Moscow 117259, Russia}
\affiliation{University of New Mexico, Albuquerque, New Mexico 87131, USA}
\affiliation{The Ohio State University, Columbus, Ohio 43210, USA}
\affiliation{Okayama University, Okayama 700-8530, Japan}
\affiliation{Osaka City University, Osaka 588, Japan}
\affiliation{University of Oxford, Oxford OX1 3RH, United Kingdom}
\affiliation{Istituto Nazionale di Fisica Nucleare, Sezione di Padova-Trento, $^{ff}$University of Padova, I-35131 Padova, Italy}
\affiliation{University of Pennsylvania, Philadelphia, Pennsylvania 19104, USA}
\affiliation{Istituto Nazionale di Fisica Nucleare Pisa, $^{gg}$University of Pisa, $^{hh}$University of Siena and $^{ii}$Scuola Normale Superiore, I-56127 Pisa, Italy}
\affiliation{University of Pittsburgh, Pittsburgh, Pennsylvania 15260, USA}
\affiliation{Purdue University, West Lafayette, Indiana 47907, USA}
\affiliation{University of Rochester, Rochester, New York 14627, USA}
\affiliation{The Rockefeller University, New York, New York 10065, USA}
\affiliation{Istituto Nazionale di Fisica Nucleare, Sezione di Roma 1, $^{jj}$Sapienza Universit\`{a} di Roma, I-00185 Roma, Italy}
\affiliation{Rutgers University, Piscataway, New Jersey 08855, USA}
\affiliation{Texas A\&M University, College Station, Texas 77843, USA}
\affiliation{Istituto Nazionale di Fisica Nucleare Trieste/Udine, I-34100 Trieste, $^{kk}$University of Udine, I-33100 Udine, Italy}
\affiliation{University of Tsukuba, Tsukuba, Ibaraki 305, Japan}
\affiliation{Tufts University, Medford, Massachusetts 02155, USA}
\affiliation{University of Virginia, Charlottesville, Virginia 22906, USA}
\affiliation{Waseda University, Tokyo 169, Japan}
\affiliation{Wayne State University, Detroit, Michigan 48201, USA}
\affiliation{University of Wisconsin, Madison, Wisconsin 53706, USA}
\affiliation{Yale University, New Haven, Connecticut 06520, USA}

\author{T.~Aaltonen}
\affiliation{Division of High Energy Physics, Department of Physics, University of Helsinki and Helsinki Institute of Physics, FIN-00014, Helsinki, Finland}
\author{B.~\'{A}lvarez~Gonz\'{a}lez$^z$}
\affiliation{Instituto de Fisica de Cantabria, CSIC-University of Cantabria, 39005 Santander, Spain}
\author{S.~Amerio}
\affiliation{Istituto Nazionale di Fisica Nucleare, Sezione di Padova-Trento, $^{ff}$University of Padova, I-35131 Padova, Italy}
\author{D.~Amidei}
\affiliation{University of Michigan, Ann Arbor, Michigan 48109, USA}
\author{A.~Anastassov$^x$}
\affiliation{Fermi National Accelerator Laboratory, Batavia, Illinois 60510, USA}
\author{A.~Annovi}
\affiliation{Laboratori Nazionali di Frascati, Istituto Nazionale di Fisica Nucleare, I-00044 Frascati, Italy}
\author{J.~Antos}
\affiliation{Comenius University, 842 48 Bratislava, Slovakia; Institute of Experimental Physics, 040 01 Kosice, Slovakia}
\author{G.~Apollinari}
\affiliation{Fermi National Accelerator Laboratory, Batavia, Illinois 60510, USA}
\author{J.A.~Appel}
\affiliation{Fermi National Accelerator Laboratory, Batavia, Illinois 60510, USA}
\author{T.~Arisawa}
\affiliation{Waseda University, Tokyo 169, Japan}
% ...

\collaboration{CDF Collaboration\footnote{With visitors from
$^a$Istituto Nazionale di Fisica Nucleare, Sezione di Cagliari, 09042 Monserrato (Cagliari), Italy,
$^b$University of CA Irvine, Irvine, CA 92697, USA,
$^c$University of CA Santa Barbara, Santa Barbara, CA 93106, USA,
$^d$University of CA Santa Cruz, Santa Cruz, CA 95064, USA,
$^e$Institute of Physics, Academy of Sciences of the Czech Republic, Czech Republic,
$^f$CERN, CH-1211 Geneva, Switzerland,
$^g$Cornell University, Ithaca, NY 14853, USA,
$^h$University of Cyprus, Nicosia CY-1678, Cyprus,
$^i$Office of Science, U.S. Department of Energy, Washington, DC 20585, USA,
$^j$University College Dublin, Dublin 4, Ireland,
$^k$ETH, 8092 Zurich, Switzerland,
$^l$University of Fukui, Fukui City, Fukui Prefecture, Japan 910-0017,
$^m$Universidad Iberoamericana, Mexico D.F., Mexico,
$^n$University of Iowa, Iowa City, IA 52242, USA,
$^o$Kinki University, Higashi-Osaka City, Japan 577-8502,
$^p$Kansas State University, Manhattan, KS 66506, USA,
$^q$Korea University, Seoul, 136-713, Korea,
$^r$UniverswU
$^s$Queen Mary, University of London, London, E1 4NS, United Kingdom,
$^t$University of Melbourne, Victoria 3010, Australia,
$^u$Muons, Inc., Batavia, IL 60510, USA,
$^v$Nagasaki Institute of Applied Science, Nagasaki, Japan,
$^w$National Research Nuclear University, Moscow, Russia,
$^x$Northwestern University, Evanston, IL 60208, USA,
$^y$University of Notre Dame, Notre Dame, IN 46556, USA,
$^z$Universidad de Oviedo, E-33007 Oviedo, Spain,
$^{aa}$CNRS-IN2P3, Paris, F-75205 France,
%
%
$^{ab}$Universidad Aut\'onoma de Madrid, Cantoblanco, 28049, Madrid,
$^{bb}$Texas Tech University, Lubbock, TX 79609, USA,
$^{cc}$Universidad Tecnica Federico Santa Maria, 110v Valparaiso, Chile,
$^{dd}$Yarmouk University, Irbid 211-63, Jordan,
}}
\noaffiliation

\noaffiliation
\begin{abstract}
      Using data from \(\proton\antiproton\) collisions at
      \(\sqrt{s}=1.96\tev\) recorded by the \(\cdf2 \) detector at the
      Fermilab Tevatron, we present improved measurements of the masses
      and first measurements of natural widths of the four bottom baryon
      resonance states \Sigbp,~\Sigbstp and \Sigbm,~\Sigbstm. These
      states are fully reconstructed in their decay modes to
      \(\Lb\pipm\) where \(\Lb\to\Lc\pim\) with \(\Lc\to\pKpi\).  The
      analysis is based on a data sample corresponding to an integrated
      luminosity of \({6.0}\invfb \) collected by an online event
      selection based on tracks displaced from the
      \(\proton\antiproton\) interaction point.
\end{abstract}
\maketitle 
%
%
\section{Introduction}
\label{sec:Intro}
  Baryons with a heavy quark \( Q \) as the ``nucleus'' and a light
  diquark \( q_{1}q_{2} \) as the two orbiting ``electrons'' can be
  viewed as the ``helium atoms'' of quantum chromodynamics (QCD).  The
  heavy quark in the baryon may be used as a probe of confinement that
  allows the study of non-perturbative QCD in a different regime from
  that of the light baryons.
\par
  Remarkable achievements in the theory of heavy quark hadrons were made
  when it was realized that a single heavy quark \(Q\) with mass 
  \( m_{Q}\gg\LQCD \) in the heavy hadron \(H_{Q} \) can be considered as a
  static color source in the hadron's rest frame~\cite{Isgur:1989vq}.
  Based on this conjecture, the light diquark properties of the charm
  baryon \( \Lc\,(\Sigc) \) and its bottom partner \( \Lb\,(\Sigb) \)
  can be related by an approximate \( SU(2) \) symmetry with
  \( \c\leftrightarrow\b \) quark exchange. Another symmetry emerges
  because the spin of the heavy quark \( S_{Q} \) decouples from the
  gluon field.
  Models exploiting these heavy quark symmetries are collectively
  identified as heavy quark effective theories
  (HQET)~\cite{Neubert:1993mb,Manohar:2000dt}.
\par 
  As the spin \( S_{qq} \) of a light diquark (plus a gluon field) and the
  spin \( S_{Q} \) of a heavy quark are decoupled in HQET, heavy baryons
  can be described by the quantum numbers
  \( S_{Q},\,m_{Q},\,S_{qq},\,m_{qq} \).  The total spins of the \(S\)-wave
  (no orbital excitation) baryon multiplets can be expressed as the sum
  \(\vec{J}\,=\,\vec{S}_{Q}\,+\,\vec{S}_{qq} \).  Then the singlet \Lb
  baryon, with quark content \( \b[\u\d] \) according to HQET, has spin
  of the heavy quark \( S_{b}^{P}={\frac{1}{2}}^{+} \).  Its flavor
  antisymmetric \( [\u\d] \) diquark has spin
  \(S_{[\u\d]}^{P}=0^{+}\)~\cite{Korner:1994nh}. Under these conditions the \b
  quark and the \( [\u\d] \) diquark make the lowest-lying singlet
  ground state \( J^{P} = {\frac{1}{2}}^{+} \).  The partner of the \Lb
  baryon in the strange quark sector is the \Lz baryon.
%
  The other two states \Sigb and \Sigbst with quark content and spin of
  the flavor symmetric \( \{qq\} \) diquark \(S_{\{qq\}}=1^{+} \),
  constitute two isospin \( I=1 \) triplets with total spin 
  \( J^{P} = {\frac{1}{2}}^{+} \) and 
  \( J^{P} = {\frac{3}{2}}^{+} \)~\cite{Korner:1994nh}. 
  These states are the lowest-lying \(S\)-wave states that can decay to
  the singlet \Lb via strong processes involving soft pion emission --
  provided sufficient phase space is available.  The \Sigb and \Sigbst
  particles are classified as bottom baryon resonant states. The
  partners of the \Sgbst states~\cite{notation:sgbst}
  in the strange quark sector are \( {\Sigs}^{(*)} \) baryon resonances,
  though the \( J^{P} = {\frac{1}{2}}^{+} \) \( {\Sigs} \) states are
  light enough to decay only weakly or radiatively, and only the 
  \(J^{P} = {\frac{3}{2}}^{+} \) states \( {\Sigs(1385)} \) decay strongly
  via the \( \Lz\pi \) mode~\cite{Nakamura:2010zzi}.
\par 
  Some recent HQET calculations for bottom baryons are available
  in~Ref.~\cite{Ebert:2005xj}.  The mass spectra of single heavy quark
  baryons calculated with HQET in combined expansions in \(1/{{m}_{Q}}\)
  and \( 1/{{N}_{c}} \), with \({{N}_{c}} \) defined as a number of colors,
  are presented in~Ref.~\cite{Jenkins:1996de}.
  In the potential quark model, the mass differences 
  \( m({\Sigma}_{Q}) - m({\Lambda}_{Q}) \) and 
  \( m({\Sigma}^{*}_{Q}) - m({\Sigma}_{Q}) \)
  are largely due to hyperfine splittings, hence the mass differences
  scale as \( 1/{{m}_{Q}} \).  Some recent predictions based on
  potential quark models are found
  in~Refs.~\cite{Rosner:2006jz,Rosner:2006yk}.
  There are striking patterns in the masses and mass differences of
  known hadrons.  Some of these regularities can be understood from
  known general properties of the interactions of quarks, without
  specifying the explicit form of the Hamiltonian.  Following this
  approach, the authors of~Ref.~\cite{Roncaglia:1994ex} use
  semi-empirical mass formulae to predict the spectra of \c and \b
  baryons.
  The non-perturbative formalism of QCD sum rules has been applied
  within HQET to calculate the mass spectra of the heavy baryons
  \( \Lambda_{Q} \) and \( \Sigma_{Q} \)~\cite{Liu:2007fg}.
  Lattice non-relativistic QCD calculations for bottom
  baryons~\cite{Mathur:2002ce} have been quite successful, though the
  uncertainties are typically large and
  exceed the uncertainties of the experimental measurements.
\par
  The mass splittings between members of the \(I=1\) isospin triplets
  \( \Sgbst \) arise from a combination of the intrinsic quark mass
  difference \( m(\d)>m(\u) \) and the electromagnetic interactions
  between quarks~\cite{Rosner:2006yk,Isgur:1979ed}.  Because of electromagnetic
  effects and the \d quark being heavier than the \u quark, the
  \( \Sgbstm \) states (with composition \( \b\{\d\d\} \) {\it i.e.} all
  quarks with negative electric charge) are expected to be heavier than
  the \( \Sgbstp \) states whose composition is
  \(\b\{\u\u\}\)~\cite{Chan:1985ty}.  No previous experimental
  measurements of isospin mass splitting of bottom baryons are
  available.
\par  
  The description of strong decays of baryon resonances is a difficult
  theoretical task~\cite{Capstick:2000qj}. Only a few
  calculations~\cite{Korner:1994nh,Guo:2007qu,Hwang:2006df} 
  of the \(\Sgbst\) natural widths are available.
  The  widths are predicted in the range 
  \(4.5 - 13.5\mevcc \) for \(\Gamma(\Sigb,{\frac{1}{2}}^{+})\), 
  and the range \( 8.5 - 18.0\mevcc \) for 
  \(\Gamma(\Sigbst,{\frac{3}{2}}^{+})\). 
\par
  Until recently, direct observation of \b baryons has been limited to the
  \(\Lb\) reconstructed in its weak decays to \(\jpsi\Lz \) and
  \( \Lc\pim \)~\cite{Nakamura:2010zzi}.  The substantially enlarged
  experimental data sets delivered by the Tevatron allow significant 
  advances in the spectroscopy of heavy quark baryon states. The
  resonance \(\Sgbst\) states were discovered by CDF~\cite{:2007rw}.
  The charged bottom strange  \(\Xi_{b}^{-}\) baryon was
  observed and measured~\cite{:2007ub,:2007un,Aaltonen:2009ny} by both
  the CDF and \dzero Collaborations. Later, \dzero reported the first
  observation of the bottom doubly-strange particle
  \(\Omega_{b}^{-}\)~\cite{Abazov:2008qm}. Subsequently the CDF
  Collaboration confirmed the signal and measured the mass of the
  \(\Omega_{b}^{-}\) baryon~\cite{Aaltonen:2009ny}.  Lastly, the neutral
  partner of \(\Xi_{b}^{-}\), the bottom strange baryon \(\Xi_{b}^{0}\),
  was reported for the first time by CDF~\cite{Aaltonen:2011wd}.
  Precise measurements of the masses and natural widths of baryon
  resonances in the charm sector, specifically the \(\Sgcstz\),
  \(\Sgcstpp\), and \(\LcS\), were recently reported by the CDF
  Collaboration~\cite{Aaltonen:2011sf}.
\par 
  This study follows the first observation of the \Sgbst
  states using \(1.1\invfb\)~\cite{:2007rw}.
  We confirm the observation of those states using a larger data sample,
  improve the measurement technique, and add new measurements of
  properties of the \Sgbst resonances. In the present analysis, the
  masses of the \Sgbstp and \Sgbstm states are determined independently,
  with no input from theory assumptions, differing from the previous CDF
  analysis~\cite{:2007rw}.
%
%
  Using an enlarged data sample of \( 6\invfb \), we extract the
  direct mass measurements with smaller statistical and systematic
  uncertainties than previously. First measurements of the natural
  widths of the \( J^{P} = {\frac{3}{2}}^{+} \) and \( J^{P} =
  {\frac{1}{2}}^{+} \) states are presented.  Based on the new mass
  measurements, we determine the isospin mass splitting for the \Sigb
  and \Sigbst isospin \( I=1 \) triplets. 
\par
  Section~\ref{sec:detector} provides a brief description of the \(\cdf2 \)
  detector, the online event selection (trigger) important for this
  analysis, and the detector simulation.  In Sec.~\ref{sec:Data} the
  data selection, analysis requirements, and reconstruction of the
  signal candidates are described. Section~\ref{sec:fitter} discusses the
  fit model of the final spectra and summarizes the fit results. In
  Sec.~\ref{sec:significance} we estimate the significance of signals
  extracted from the fits.  The systematic uncertainties are discussed
  in Sec.~\ref{sec:sys}.  
  We present a summary of the measurements and conclusions in
  Sec.~\ref{sec:results}.
\section{The CDF~II detector and simulation}
\label{sec:detector}
  The component of the \(\cdf2 \) detector~\cite{Acosta:2004yw} most
  relevant to this analysis is the charged particle tracking
  system. The tracking system operates in a uniform axial magnetic
  field of \(1.4\,{\rm T}\) generated by a superconducting solenoidal magnet.
\par 
  The \(\cdf2 \) detector uses a cylindrical coordinate system with
  \(z\) axis along the nominal proton beam line, radius \(r\) measured
  from the beam line and \(\phi\) defined as an azimuthal angle. The
  transverse plane \((r,\phi)\) is perpendicular to the \(z\) axis.  The
  polar angle, \(\theta\), is measured from the \(z\) axis.  The
  impact parameter of a charged particle track \(d_{0}\) is defined as
  the distance of closest approach of the particle track to the primary
  vertex in the transverse plane. Transverse momentum, \(\pt\), is the
  component of the particle's momentum projected onto the transverse
  plane.  Pseudorapidity is defined as 
  \(\eta\,\equiv\,-\ln(\tan(\theta/2))\).
\par 
  The inner tracking system comprises three silicon detectors:
  layer~00~(L00), {the silicon vertex detector}~(SVX~II) and {the intermediate
  silicon layers}~(ISL)~\cite{Sill:2000zz,Affolder:2000tj,Nahn:2003tm,Hill:2004qb}. 
  The innermost part, the L00 detector, is a layer of single-sided
  radiation tolerant silicon sensors mounted directly on the beam pipe
  at a radius of \(1.35-1.6\cm\) from the proton beam line.  It provides
  only an \({r}\)-\({\phi}\) measurement and enhances the
  impact parameter resolution.
  Outside this, the five double-sided layers of SVX~II provide up to 10
  track position measurements.
  Each of the layers provides an \({r}\)-\({\phi}\) measurement, while three
  return a measurement along \(z\), and the other two return a
  measurement along a direction oriented at \(\pm1.2\degrees\) to the
  \(z\) axis.
  The SVX~II spans the radii between \(2.5\cm\) and \(10.6\cm\) and
  covers the pseudorapidity range \(\rapid <2.0\). The SVX~II
  detector provides a vertex resolution of approximately
  \(15\mkm \) in the transverse plane and \(70\mkm \)
  along the \(z\) axis. 
  A fine track impact parameter resolution
  \({\sigma_{d_{0}}}\simeq{35}\mkm\) is achieved, where the
  \({\sigma_{d_{0}}}\) includes an approximate \({28}\mkm \)
  contribution from the actual transverse size of the beam spot. 
  The outermost silicon subdetector, ISL, consists of double-sided
  layers at radii \(20\cm\) to \(28\cm\), providing two or four
  hits per track depending on the track pseudorapidity within 
  the range \(\rapid <2.0\) instrumented by the ISL. 
\par  
  A large open cell cylindrical drift chamber, the central outer
  tracker~(COT)~\cite{Affolder:2003ep}, completes the CDF
  detector tracking system.   The COT consists of \(96\) sense wire
  layers arranged in \(8\) superlayers of \(12\) wires each.  Four of
  these superlayers provide axial measurements, and four provide stereo
  views at \(\pm2\degrees\). 
  The active volume of the COT spans the radial region from \(43.4\cm\)
  to \(132.3\cm\). The pseudorapidity range \(\rapid <1.0\) is covered
  for tracks passing through all layers of the COT, while for the range
  out to \(1.0<\rapid<2.0\), tracks pass through less than the full
  \(96\) layers.
%
%
  The trajectory of COT tracks is extrapolated into the SVX~II detector,
  and the tracks are refitted with additional silicon hits consistent
  with the track extrapolation. 
  The two additional layers of the ISL
  help to link tracks in the COT to hits in the SVX~II.
  The combined track transverse momentum resolution is 
  \( {\sigma({\pt})}/{\pt}\simeq{0.07\%}\,{\pt}\,[\gevc]^{-1} \).
\par  
  The analysis presented here is based on events recorded with a
  three-tiered trigger system configured to collect large data samples
  of heavy hadrons decaying through multi-body hadronic channels.  We
  refer to this as the displaced two-track trigger.  We use two
  configurations of this trigger, the ``low-\pt'' and the
  ``medium-\pt'' selections.  At level~1, the trigger uses
  information from the hardware extremely fast
  tracker~\cite{Thomson:2002xp}. The ``low-\pt'' configuration of
  the displaced two-track trigger requires two tracks in the COT with 
  \(\pt >2.0\gevc \) for each track, and with an opening angle of
  \(\left|\Delta\phi\right| <90\degrees \) between the tracks in the
  transverse plane. Additionally the track pair scalar sum must satisfy
  \({\pt}_{1}+{\pt}_{2} >4.0\gevc\). The corresponding criteria imposed
  in the ``medium-\pt'' configuration are \( \pt >2.0\gevc \) for each
  track, opening angle \(\left|\Delta\phi\right|<135\degrees \), and
  \({\pt}_{1}+{\pt}_{2}>5.5\gevc\).
  The level~2 silicon vertex trigger~(SVT)~\cite{Ashmanskas:2003gf,RistoriPunzi:CDFTrigger}
  associates the track pair from the extremely fast tracker with hits in
  the SVX~II detector and recognizes both tracks using a large look-up
  table of hit patterns. The SVT repeats the level~1 \pt criteria and
  limits the opening angle to \( 2\degrees< \left|\Delta\phi\right| <90\degrees \).
  Only in the case of the medium-\pt configuration are
  the charges of the tracks required to be of opposite sign. Crucially,
  the SVT imposes a requirement on the transverse impact parameter of
  each track to be \(0.12\,<{d_{0}}\,<1\mm\), given the excellent
  resolution provided by SVX~II. 
%
%
  Finally, the distance in the transverse plane between the beam axis
  and the intersection point of the two tracks projected onto their
  total transverse momentum is required to be \(\lxy>200\mum\).
  The level~3 software trigger uses a full reconstruction of the event
  with all detector information and confirms the criteria applied at
  level~2.  The trigger criteria applied to the \({d_{0}}\) of each
  track in the pair and to \(\lxy \) preferentially select decays of
  long-lived heavy hadrons over prompt background, ensuring that the
  data sample is enriched with \b hadrons.
\par 
  The mass resolution on the \Sgbst resonances is predicted with a
  Monte Carlo simulation that generates \b quarks according to a
  next-to-leading order calculation~\cite{Nason:1987xz} and produces
  events containing final state hadrons by simulating \b quark
  fragmentation~\cite{Peterson:1982ak}.  Mass values of \( 5807.8\mevcc \) 
  for \Sigb and \( 5829.0\mevcc \) for \Sigbst~\cite{:2007rw} are used
  in the Monte Carlo generator.  Final state
  decay processes are simulated with the {\sc evtgen}~\cite{Lange:2001uf} 
  program, and all simulated \b hadrons are
  produced without polarization.  The generated events are input to the
  detector and trigger simulation based on {\sc geant3}~\cite{Geant} and
  processed through the same reconstruction and analysis algorithms as
  are used on the data.
\section{Data sample and event selection}
\label{sec:Data}
  This analysis is based on data equivalent to \(6.0~\invfb\) of \(\proton\antiproton\)
  collisions collected with the displaced two-track trigger
%
%
  between March 2002 and February 2010.  We study \Sgbst resonances in
  the exclusive strong decay mode
  \(\Sgbstpm\to\Lb\pipm_{\mathit{s}}\), where 
  the low momentum pion \(\pipm_{\mathit{s}} \) is produced near 
  kinematic threshold~\cite{notation:cc}.
  The \( \Lb \) decays to \( \Lc\pim_{b} \) with a prompt pion
  \(\pim_{b}\) produced in the weak decay. This is followed by the weak decay
  \(\Lc\to\pKpi\).
\par 
  To reconstruct the parent baryons, the tracks of charged particles are
  combined in a kinematic fit to form candidates. No particle identification is
  used in this analysis. The following two complementary
  quantities defined in the plane transverse to the beam line and
  relating the decay path of baryons to their points of origin are used:
  the proper decay time of the baryon candidate \(h\) expressed in
  length units \( \ct{(h)} \), and the impact parameter \( d_{0}{(h)} \).
  Specifically, the decay length is defined as
  \begin{equation}  
    \ct{(h)} = \lxy{(h)}\,\frac{M{(h)}\,{c}}{\pt{(h)}}\,\,,
  \label{eq:ctau}
  \end{equation}   
  where \( \lxy{(h)} \) is expressed in length units and defined as the
  projection onto \( {\vec{p}}_{\rm T}{(h)} \) of the vector connecting the
  primary vertex to the heavy baryon decay vertex in the transverse
  plane. 
%
%
  The transverse impact parameter \( d_{0}{(h)} \) of the candidate is
  defined analogous to the one of a charged particle track.  An
  event-specific primary interaction vertex is used in the calculation
  of the \( \ct{(h)} \) and \( d_{0}{(h)} \) quantities.  The
  measurement uncertainties \( \sigma_{\it{ct}} \) and \( \sigma_{d_{0}} \)
  originate from the track parameter uncertainties and the uncertainty
  on the primary vertex.
\subsection{Reconstruction of the \Lb candidates}
\label{sec:reco-lb}
  The analysis begins with reconstruction of the \(\LcpKpi\) decay
  by fitting three tracks to a common vertex.  
  The invariant mass of the \(\Lc\) candidate is required to be within
  \(\pm18\mevcc \) of the world-average \(\Lc\)
  mass~\cite{Nakamura:2010zzi}.
  The momentum vector of the \(\Lc\) candidate is then extrapolated to
  intersect with a fourth pion track, the \(\pim_{b}\)-candidate, to
  form the \( \Lb\to\Lc\pim_{b} \) candidate vertex. The \Lb vertex is
  subjected to a three-dimensional kinematic fit with the \(\Lc\)
  candidate mass constrained to its world average
  value~\cite{Nakamura:2010zzi}. The probability of the constrained \Lb
  vertex fit must exceed \( 0.01\% \).
  Standard quality requirements are applied to each track, and only
  tracks with \(\pt>400\mevc \) are used. 
  All tracks are refitted using pion, kaon and proton mass hypotheses to
  properly correct for the differences in multiple scattering and
  ionization energy loss.
  At least two tracks among the \(\proton,\,\Km,\,\pip\), and
  \(\pim_{b}\) candidates are required to fulfill the level~2~(SVT)
  trigger requirements.
%
\par 
  To suppress prompt backgrounds from the primary interaction, the decay
  vertex of the \Lb is required to be distinct from the primary vertex.
  To achieve this, cuts on \( \ct{(\Lb)} \) and its significance
  \(\ct(\Lb)/\sigma_{\it{ct}} \) are applied.  We require the \Lc vertex
  to be close to the \Lb vertex by applying cuts on  
%
  \(\ct(\Lc)\) where the corresponding quantity \( \lxy(\Lc) \) is
  calculated with respect to the \Lb vertex. The requirement
  \(\ct(\Lc)>-150\,\mkm \) reduces contributions from \Lc baryons
  directly produced in \(\proton\antiproton\) interaction and from
  random combination of tracks faking \Lc candidates which may have
  negative \(\ct(\Lc)\) values. The other restriction,
  \(\ct(\Lc)<250\,\mkm \), aims at reducing contributions from
  \(\Bdb\to\Dp\pim\) decays, followed by \(\Dp\to\Km\pip\pip\) decays. The
  requirements take into account \(\ct\) resolution effects and exploit
  the much shorter \Lc lifetime compared
  to the \Dp~\cite{:2007rw,Abulencia:2006df}.
  To reduce combinatorial background and contributions from partially
  reconstructed decays,
  we ask \Lb candidates to point to the primary vertex by requiring the
  impact parameter \({{d_{0}}(\Lb)}\) not to exceed \( 80\,\mkm \).
%
%
  The choice of analysis requirements to identify \(\Lb\to\Lc\pim_{b}\)
  candidates is made using an optimization based on the experimental
  data only.  The figure of merit \( S/\sqrt{S+B} \) is used during the
  optimization, where \(S\) is the \Lb signal and \(B\) is the
  background under the signal, respectively. At every step of the
  optimization procedure, both quantities are obtained from fits of the
  \( \Lc\pim_{b} \) invariant mass spectrum and are determined from the
  corresponding numbers of candidates fit within \( \pm3\sigma \)
  of the \Lb signal peak.
  Table~\ref{tab:lb-cuts} summarizes the resulting \Lb analysis requirements.
%
\par
\begin{figure}
\begin{center}
  \includegraphics[width=0.5\textwidth]
  {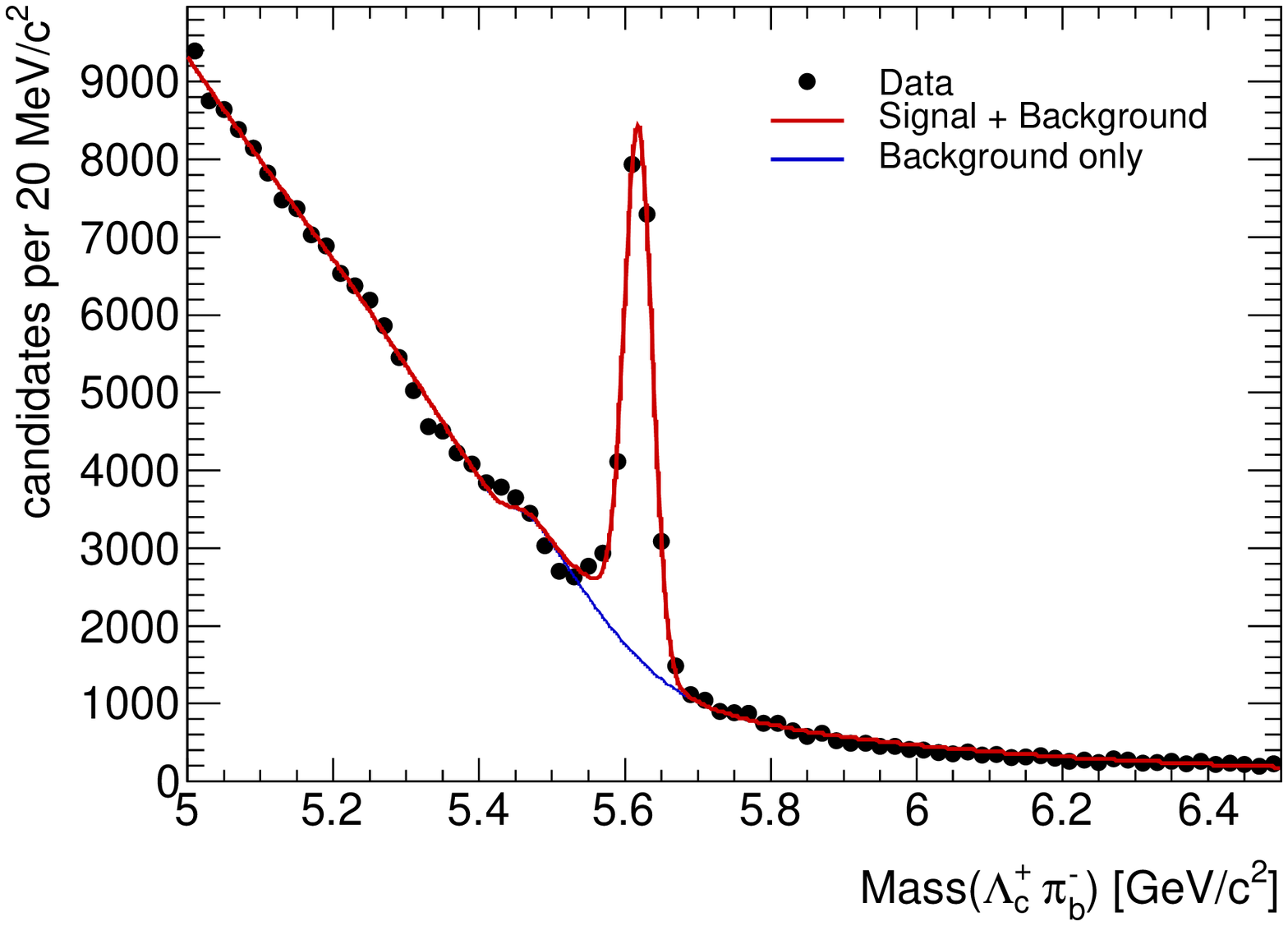}
\caption{ Invariant mass distribution of \(\Lb\to\Lc\pim_{b}\)
          candidates with the projection of a mass fit overlaid.
          \label{fig:signal-lb} }
\end{center}
\end{figure}

 
   Figure~\ref{fig:signal-lb} shows a prominent \Lb signal in the
   \( \Lc\pim_{b} \) invariant mass distribution, reconstructed using
   the optimized criteria.  A binned maximum-likelihood fit finds a
   signal of approximately \( 16\,300 \) candidates at the expected \Lb
   mass, with a signal to background ratio around \(1.8\).
  The fit model describing the invariant mass distribution comprises the
  Gaussian \(\Lb\to\Lc\pim_{b}\) signal on top of a background shaped by
  several contributions. Random four-track combinations dominating
  the right sideband are modeled with an exponentially decreasing
  function.  Coherent sources populate the left sideband and leak under
  the signal. These include reconstructed \B mesons that pass the
  \(\Lb\to\Lc\pim_{b}\) selection criteria, partially reconstructed \Lb
  decays, and fully reconstructed \Lb decays other than \(\Lc\pim_{b}\)
  (\exegrat~\(\Lb\to\Lc\Km\)). Shapes representing
  the physical background sources are derived 
  from Monte Carlo simulations.  Their normalizations are constrained to
  branching ratios that are either measured (for \B meson
  decays, reconstructed within the same \(\Lc\pim_{b}\) sample) or
  theoretically predicted  
  (for \Lb decays)~\cite{:2007rw,Abulencia:2006df}.
\begin{table}
  \begin{center}
  \caption{ Analysis requirements for \(\Lb\to\Lc\pim_{b}\)
            reconstruction. The quantity \( \ct(\Lc\leftarrow\Lb) \) is
            defined analogously to Eq.~(\ref{eq:ctau}) as the \( \Lc \)
            proper time where \( \lxy(\Lc) \) is calculated with respect
            to the \( \Lb \) vertex. 
            \label{tab:lb-cuts} }
  \begin{tabular}{ll}
  \hline
  \hline
   Quantity  &  Requirement   \\
  \hline
    \(\ct(\Lb) \)                    &  \( >200\,\mkm  \) \\
    \(\ct(\Lb)/\sigma_{\it{ct}} \)       &  \( >12.0  \) \\
    \({{d_{0}}(\Lb)}\)               &  \( <80\,\mkm \) \\
    \(\ct(\Lc \leftarrow \Lb) \)     &  \( >-150\,\mkm \) \\
    \(\ct(\Lc \leftarrow \Lb) \)     &  \( <250\,\mkm \) \\
    \(\pt(\pim_{b})\)                &  \( >1.5\,\gevc \) \\
    \(\pt(\Lb)\)                     &  \( >4.0\,\gevc \) \\
    \({\rm Prob}(\chi^{2}_{3D})\) of \(\Lb\) vertex fit & \(>0.01\%\) \\
  \hline
  \hline
  \end{tabular}
  \end{center}
\end{table}
\subsection{Reconstruction of \Sgbstpm candidates}
\label{sec:reco-sgb}
  To reconstruct the \(\Sgbstpm\to\Lb\pipm_{\mathit{s}}\) candidates,
  each \(\Lc\pim_{b}\) candidate with invariant mass within the \Lb
  signal region,
  \(5.561-5.677\,\gevcc\),
  is combined with one of the tracks remaining in the event with
  transverse momentum down to \(200\mevc \).
%
  The \Lb mass range covers \( \pm3 \) standard deviations as determined
  by a fit to the signal peak of Fig.~\ref{fig:signal-lb}.
  To increase the efficiency for reconstructing \Sgbstpm decays near the
  kinematic threshold, the quality criteria applied to soft pion tracks
  are loosened in comparison with tracks used for the \Lb candidates.  The
  basic COT and SVX~II hit requirements are imposed on
  \(\pipm_{\mathit{s}}\) tracks, and only tracks with a valid track fit
  and error matrix are accepted.
\par  
  Random combinations of \Lb signal candidates with
  \(\pipm_{\mathit{s}}\) tracks 
  constitute the dominant background to the
  \(\Sgbstpm\to\Lb\pipm_{\mathit{s}}\) signal. The remaining
  backgrounds are random combinations of soft tracks with \B
  mesons reconstructed as \Lb baryons, and combinatorial background
  events~\cite{:2007rw}. 
  To reduce the background level, a kinematic fit is applied to the
  resulting combinations of \(\Lb \) candidates and soft pion tracks
  \(\pipm_{\mathit{s}}\) to constrain them to originate from a common
  point.  Furthermore, since the bottom baryon resonance originates and
  decays at the primary vertex, the soft pion track is required to point
  back to the primary vertex by requiring an impact parameter
  significance,
   \( {{d_{0}}(\pipm_{\mathit{s}})/{\sigma_{d_{0}}}} \), 
  smaller than three. The transverse momentum of the soft pion is
  required to be smaller than the \(\pim_{b}\) transverse momentum.  As
  we already require \( \pt(\pim_{b})>\,1.5\gevc \)
  (Table~\ref{tab:lb-cuts}) the condition imposed on the soft pion \(\pt\)
  is fully efficient. The \Sgbstpm candidate selection
  requirements are summarized in Table~\ref{tab:sgb-cuts}.
%
%
%
\begin{table}
  \begin{center}
  \caption{ \Sgbstpm candidate selection requirements. 
           \label{tab:sgb-cuts} }
  \begin{tabular}{ll}
  \hline
  \hline 
   Quantity                                   &  Requirement \\
  \hline
  \( m(\Lc\pim_{b}) \)     & \(\in(5.561,\,5.677)\gevcc \) \\
%
  \( {{d_{0}}(\pipm_{\mathit{s}})} \)  & \( < 0.1\,\cm\ \) \\
  \( \pt(\pipm_{\mathit{s}}) \)                   & \( > 200\,\mevc \) \\
  \( {{d_{0}}(\pipm_{\mathit{s}})/{\sigma_{d_{0}}}} \) & \( < 3.0 \) \\
  \( \pt(\pipm_{\mathit{s}}) \)                           & \( <\pt(\pim_{b}) \) \\
  \(\pt(\Sgbstpm)\)                       & \(>4.0\,\gevc\) \\
  \hline
  \hline
  \end{tabular}
  \end{center}
\end{table}
%
%
\section{Determination of resonance properties}
\label{sec:fitter}
  The analysis of the \Sgbstpm mass distributions is performed using the
  \(Q\) value 
  \begin{equation} 
    Q = m(\Lb\pipm_{\mathit{s}}) - m(\Lb) - {m_{\pi}}\,\,, 
  \label{eq:qvalue}
  \end{equation} 
  where \( {m_{\pi}} \) is the known charged pion mass~\cite{Nakamura:2010zzi} 
  and \( m(\Lb) \) is the reconstructed \(\Lc\pim_{b}\) mass.
  The mass resolution of the \Lb signal and most of the systematic
  uncertainties cancel in the mass difference spectrum.
  The \( \Sigbpm \) and \( \Sigbstpm \) signals are reconstructed as two
  narrow structures in the \(Q\)-value spectrum.  The properties,
  yields, and significance of the resonance candidates are obtained by
  performing unbinned maximum-likelihood fits on the \(Q\)-value 
  spectra.
\par  
%
%
  The shapes of the \Sgbstpm resonances are each modeled with a
  non-relativistic Breit-Wigner function. Since the soft pion in
  \Sgbstpm strong decay modes is emitted in a \(P\)-wave, the width of
  the Breit-Wigner function is modified as follows~\cite{Jackson:1964zd}:
  \begin{equation} 
    \Gamma(Q;Q_{0},\Gamma_{0}) = 
    \Gamma_{0}\,{\left( \frac{p^{*}_{\pi_{\mathit{s}}}}{p^{*0}_{\pi_{\mathit{s}}}} \right)}^{3}\,,
  \label{eq:modwid}
  \end{equation} 
   where \(Q_{0}\) is the \(Q\) value at the resonance pole; 
   \(p^{*}_{\pi_{\mathit{s}}}\) and \(p^{*0}_{\pi_{\mathit{s}}}\)
   are the momenta of the soft pion in the \( \Sgbstpm \) rest frame,
   off and on the resonance pole respectively; and \( \Gamma_{0} \) is
   the corrected width. The soft pion momenta are calculated based on
   two-body decay kinematics~\cite{Nakamura:2010zzi}. Both \(Q_{0}\) and
   \( \Gamma_{0} \) are floating fit parameters.
\par 
  The Breit-Wigner function is convoluted with the detector resolution,
  which is described by a narrow core Gaussian plus a broad
  Gaussian. Their widths \( \sigma_{n} \) and \(\sigma_{w}\) and
  relative weights \( g_{n} \) and \((1-g_{n}) \) are calculated from
  the CDF full Monte Carlo simulation.
  Numerical convolution is necessary because the modified width depends
  on the mass.
  The effects of imperfect modeling in the simulation are discussed with
  the systematic uncertainties in Sec.~\ref{sec:sys}.
\par
  We use a kinematically motivated model for the background, described by
  a second order polynomial modulated with a threshold square root-like
  term,  
  \begin{equation} 
 \begin{split}
    \mathcal{BG}(Q;{m_{T}},C,b_{1},b_{2}) \,=\, & \sqrt{(Q+m_{\pi})^{2}\,-\,{m_{T}}^{2} }\,\times \\
                                                 & \mathcal{P}^{2}(Q;C,b_{1},b_{2})\,, 
 \end{split}
  \label{eq:bgr}
  \end{equation} 
  where \(C\), \(b_{1}\), and \(b_{2}\) are the second order
  \(\mathcal{P}^{2}\) polynomial coefficients and \(m_{T}\) is a
  threshold fixed to \(0.140\gevcc\), the mass of the pion.
%
%
%
\par
  The full model for the \(Q\)-value spectra of all isospin partner
  states \(\Sgbstp\) and \(\Sgbstm\) describes two narrow structures on
  top of a smooth background with a threshold.
  The negative logarithm of the extended likelihood function (NLL) is
  minimized over the unbinned set of \(Q\) values observed for 
  \(N\) candidates in data:
%
%
\begin{equation}
 \begin{split}
   -\ln{(\mathcal{L})}\,=\, & -{\sum_{k=1}^{N}}\ln( N_{1}\,{\mathcal{S}}_{1}\,
                               +\,N_{2}\,{\mathcal{S}}_{2} + N_{b}\,{\mathcal{BG}} )\\
                             & + (N_{1} + N_{2} + N_{b})\\
                             & - {N}\ln{(N_{1} + N_{2} + N_{b})}\,. 
 \end{split}
 \label{eq:lh}
\end{equation}
%
  Independent likelihood functions are used for 
  \Sgbstp and \Sgbstm candidates.
  The \(Q\)-value spectrum is fit over the range
  \(0.003-0.210\gevcc\).  
  The effect of this choice is discussed in
  Sec.~\ref{sec:sys}.
  The probability density functions (PDF) 
  in Eq.~(\ref{eq:lh}) are defined as follows:
\begin{enumerate}[(i)]
  \item \( {\mathcal{S}}_{i}\,=\,
           {\mathcal{S}}(Q;Q_{0}^{i},\Gamma_{0}^{i},\sigma_{n}^{i},g_{n}^{i},\sigma_{w}^{i}) \) 
        is the normalized convolution of a Breit-Wigner and a double
        Gaussian responsible for the \( \Sigbp\,(\Sigbm)\)~(\({i=1}\)) or 
        \(\Sigbstp\,(\Sigbstm)\)~(\({i=2}\)) signals. Here \( Q_{0}^{i} \) is
        the floating  pole mass and \( \Gamma_{0}^{i} \)
        is the floating natural width. The detector's
        Gaussian resolution parameters \(\sigma_{n}^{i},\sigma_{w}^{i}\)
        and \(g_{n}^{i}\) are set from the Monte Carlo
        data. 
        A dominant with \(g_{n}\sim70\% \) relative weight narrow core
        \(\sigma_{n}\) of about \(1.2\mevcc\) is set for the 
        \(\Sigbp\,(\Sigbm)\) and about \(1.4\mevcc \) for
         \(\Sigbstp\,(\Sigbstm)\). A broad component \(\sigma_{w}\) of
        about \(2.9\mevcc\) is set for the \( \Sigbp\,(\Sigbm)\) and
        about \(3.8\mevcc\) for \(\Sigbstp\,(\Sigbstm)\).
  \item \( N_{i} \) is the floating yield of the
        \( \Sigbp\,(\Sigbm)\)~(\({i=1}\)) 
        or \( \Sigbstp\,(\Sigbstm)\)~(\({i=2}\)).
  \item \( {\mathcal{BG}}\,=\,{\mathcal{BG}}(Q;m_{T},C,b_{1},b_{2}) \) is 
        the PDF corresponding to the background form in Eq.~(\ref{eq:bgr}). 
  \item \( N_{b} \) is the floating yield of the background
        contribution. The sum of fitted yields, 
        \(N_{1}+N_{2}+N_{b}\), is the Poisson mean value of the total
        number of candidates \( {N} \) for the particular species
        \( \Sigbp,\,\Sigbstp \) or \( \Sigbm,\,\Sigbstm \) corresponding
        to isospin triplets \( \Sigb \) and \( \Sigbst \).
\end{enumerate}
  The total number of floating parameters in the fit per each pair of
  isospin partners is nine.
\par
  Extensive tests on several thousand statistical trials show that the
  likelihood fit yields unbiased estimates with proper uncertainties.
\par
\begin{figure*}  
\begin{center} 
  \includegraphics[width=0.49\textwidth]
  {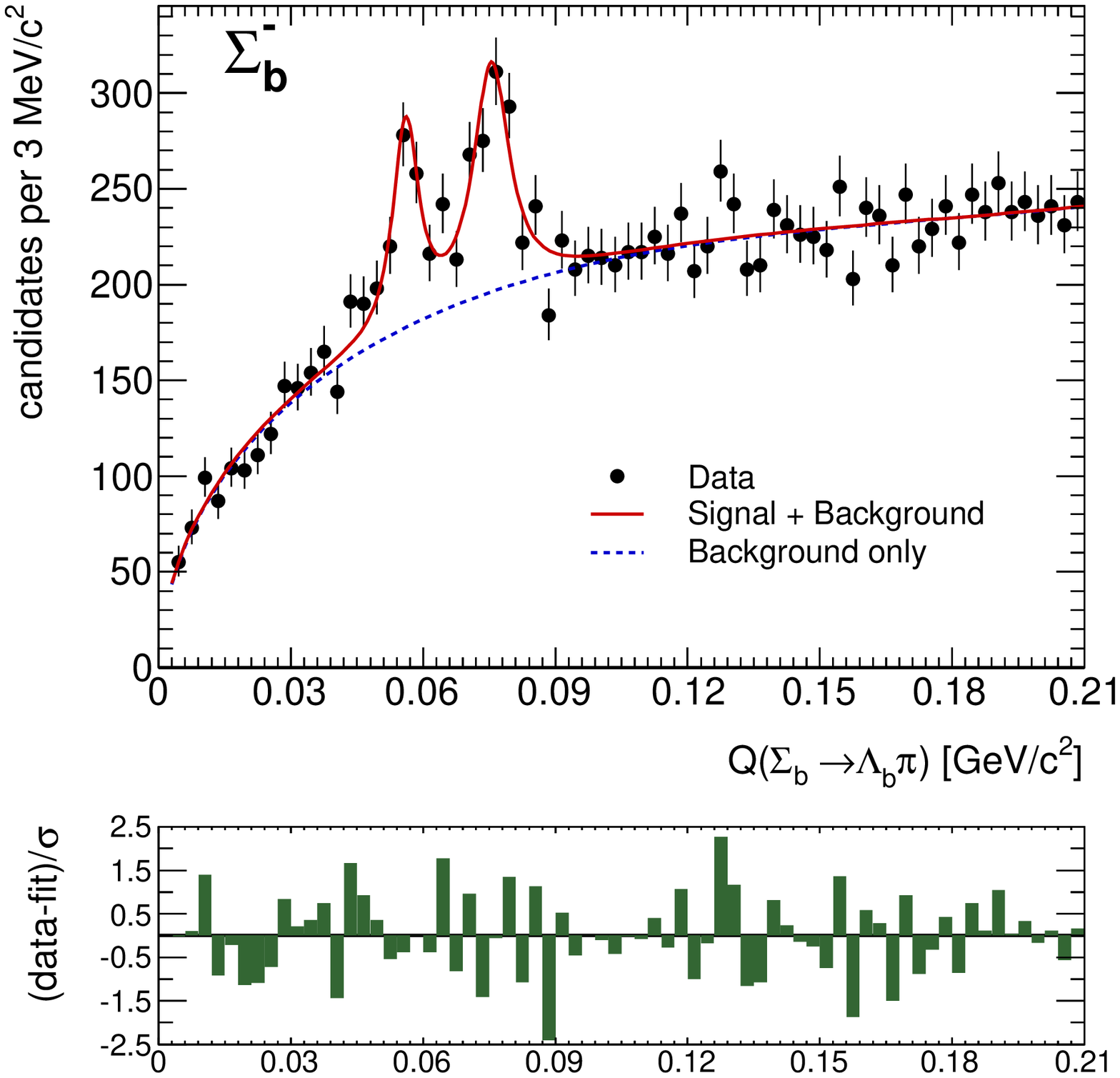}  
  \includegraphics[width=0.49\textwidth]
  {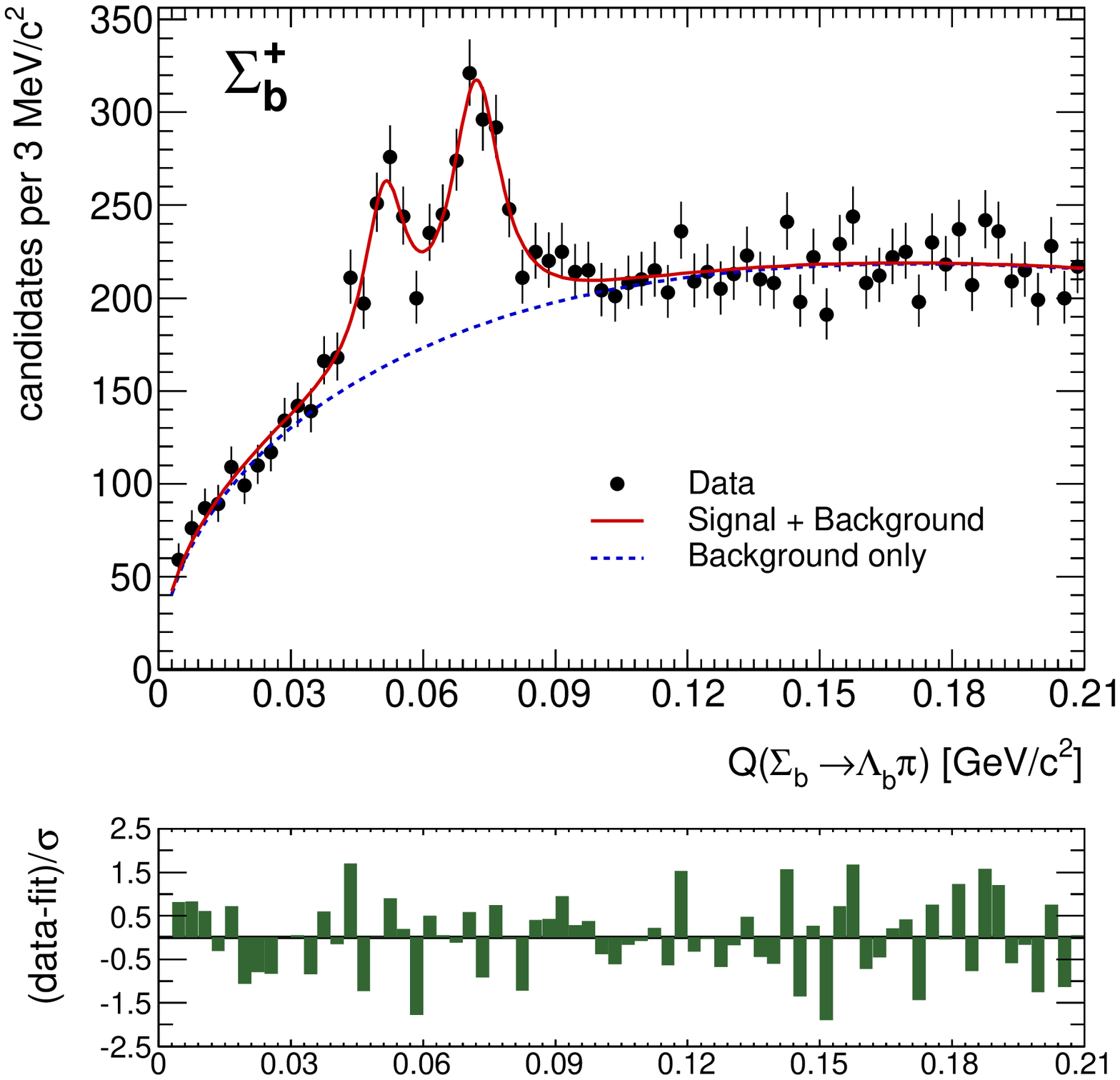}
  \caption{ The left (right) plot shows the \(Q\)-value spectrum for
            \Sgbstm (\Sgbstp) candidates with the projection of the
            corresponding unbinned likelihood fit superimposed. The
            \(Q\) value is defined in Eq.~(\ref{eq:qvalue}).  The pull
            distribution of each fit is shown in the bottom of the
            corresponding plot. }
\label{fig:signal-sgbmp}
\end{center}
\end{figure*}

  The experimental \Sgbstm and \Sgbstp \(Q\)-value distributions, each
  fitted with the unbinned likelihoods described above, are shown in
  Fig.~\ref{fig:signal-sgbmp}.  The projection of the corresponding
  likelihood fit is superimposed on each graph.
  The \(Q\)-value distributions show clear signals of
  \( \Sigbm,\,\Sigbstm \) and \( \Sigbp,\,\Sigbstp \), respectively. The
  pull distributions are shown in the bottom plots of both figures and
  are calculated as the residuals of the histogram with respect to the
  corresponding likelihood fit projection normalized by the data
  uncertainty. Both pull distributions are evenly distributed around
  zero with fluctuations of \(\pm2\sigma\), approximately.
The fit results are given in Table~\ref{tab:fitresults}.
\begin{table}[h]
\begin{center}
\caption{ Summary of the results of the fits to the 
          \({Q = M(\Lb\pipm)-M(\Lb)-m_{\pi}}\) spectra.  
          The statistical uncertainties are
          returned by the unbinned maximum-likelihood fits.
          \label{tab:fitresults} }
  \begin{tabular}{lccc}
\hline
\hline 
    State        & \(Q_{0}\) value,   & Natural width,      & Yield \\
                 & \mevcc         & \(\Gamma_{0}\), \mevcc  &  \\
\hline
    {\Sigbm}   & {\({56.2}_{-0.5}^{+0.6} \)} & {\({4.9}_{-2.1}^{+3.1} \)} & {\(340_{-70}^{+90}  \)} \\
    {\Sigbstm} & {\({75.8}\pm{0.6} \)}       & {\({7.5}_{-1.8}^{+2.2} \)} & {\(540_{-80}^{+90}  \)} \\
    {\Sigbp}   & {\({52.1}_{-0.8}^{+0.9}\)}  & {\({9.7}_{-2.8}^{+3.8} \)} & {\(470_{-90}^{+110} \)} \\
    {\Sigbstp} & {\({72.8}\pm{0.7}\)}        & {\({11.5}_{-2.2}^{+2.7}\)} & {\(800_{-100}^{+110}\)} \\
\hline
\hline
\end{tabular}
\end{center}
\end{table}
\section{Signal significance}
\label{sec:significance}
  The significance of the signals is determined using a
  \({\log}\)-likelihood ratio statistic~\cite{Wilks,Royall},
  \(\,-2\,\ln({\mathcal{L}_0}/{\mathcal{L}_1})\,\).
%
%
%
%
%
%
%
  We define hypothesis \({\mathcal{H}_1}\) corresponding to the
  presence  of \( \Sigbm,\,\Sigbstm \) or \( \Sigbp,\,\Sigbstp \)
  signals on top of the background. The \({\mathcal{H}_1}\) hypothesis
  is described by the likelihood \({\mathcal{L}_1}\); see
  Eq.~(\ref{eq:lh}).  The various null hypotheses, each identified with
  \({\mathcal{H}_0}\) and nested to \({\mathcal{H}_1}\) correspond to a
  few different less complex  scenarios described by the likelihood
  \({\mathcal{L}_0}\).
%
  The likelihood ratio  
  is used as a \(\chisq\) variable to derive
  \(p\) values for observing a deviation as large as is in our data or
  larger, assuming \({\mathcal{H}_0}\) is true. The number of degrees of
  freedom of the \(\chisq\) equals the difference \(\Delta{N_{\rm dof}}\)
  in the number degrees of freedom between the \({\mathcal{H}_1}\) and
  \({\mathcal{H}_0}\) hypotheses in each case.  We consider the
  following types of \({\mathcal{H}_0}\) to estimate the significance of
  the two-peak signal structure and of individual peaks of the observed
  \Sgbstm and \Sgbstp states:
  \begin{enumerate}[(i)]
    \item A single enhancement is observed anywhere in the fit range.
          The corresponding likelihood \({\mathcal{L}_0}\) includes only
          a single peak PDF on top of the background form in
          Eq.~(\ref{eq:bgr}), the same as for the \({\mathcal{L}_1}\).
          The difference in the number of degrees of freedom is
          \(\Delta{N_{\rm dof}} = 3\).  The width \(\Gamma_{0}\) floats
          in the fit over the wide range \(1-70\mevcc\). The position of
          the enhancement \(Q_{0}\) is allowed to be anywhere within the
          default fit range.  We test the case in which the observed two
          narrow structures could be an artifact of a wide bump where a
          few bins fluctuated down to the background level. 
%
%
    \item The signal \Sigbst is observed but the \Sigb is interpreted as
          background. We impose a loose requirement on the existence of the
          second peak, \Sigbst fixing only the width of \Sigbst to the
          expected theoretical value of \(12\mevcc\)~\cite{Guo:2007qu}.
          We let the fitter find the \Sigbst position within the default
          fit range. The number of free parameters is changed by \(4\). 
    \item The signal \Sigb is observed but the \Sigbst is interpreted as
          background.  This null hypothesis is similar to the previous
          one.  The width of the \Sigb is fixed to 
          \(7\mevcc\)~\cite{Guo:2007qu}.
    \item Neither the \Sigb nor the \Sigbst is observed, and the
          \({\mathcal{H}_0}\) hypothesis is the default background model
          used in \({\mathcal{L}_1}\).  We consider the case in which
          the smooth background fluctuates to two narrow structures
          corresponding to the \({\mathcal{H}_1}\) hypothesis. The
          difference in the number of degrees of freedom is \(6\).
  \end{enumerate}
  In addition to all the cases considered above, we introduce an
  additional case in which the \({\mathcal{H}_1}\) hypothesis
  corresponds to any single wide enhancement considered in \textrm{(i)}
  while the \({\mathcal{H}_0}\) hypothesis is the default background
  considered in \textrm{(iv)}.  This special test determines the
  significance of the single enhancement with respect to pure
  background.
\par
  Table~\ref{tab:significance} summarizes the results of these tests.
  The null hypothesis most likely to resemble our signal is a broad
  single enhancement fluctuating to the two narrow structures.  The
  results of this study establish conclusively the \Sgbstm and \Sgbstp
  signals with significance of \(6\sigma \) or higher.
\begin{table}
\begin{center}
\caption{ Statistical significances of the observed signals against
          various null hypotheses.  \(N_{\sigma}\) is the calculated
          number of Gaussian standard deviations based on
          Prob(\(\chi^2\)). }
\begin{tabular}{clcc}
\hline 
\hline
\({\mathcal{H}_0}\)    & States &  \(N_{\sigma}\) & \({\mathcal{H}_1}\) \\
\hline
Any single wide        & \Sgbstm &  \(6.7\) & Two narrow\\
enhancement            & \Sgbstp &  \(6.1\) & structures\\
                       &         &          &\\
No structures          & \Sgbstm & \(10.7\) & Any single wide\\
                       & \Sgbstp &  \(13.2\) & enhancement\\
                       &         &           &\\
No \Sigb, with \Sigbst,  & \Sgbstm &  \(7.6\) & Two narrow\\
\(\Gamma_{02}=12\mevcc\) & \Sgbstp &  \(7.9\) & structures\\
                         &         &          & \\
No \Sigbst, with \Sigb, & \Sgbstm &  \(10.0\) & Two narrow\\
\(\Gamma_{01}=7\mevcc\) & \Sgbstp &  \(12.5\) & structures\\
                        &         &           &\\
No structures          & \Sgbstm &  \(12.4\) & Two narrow\\
                       & \Sgbstp &  \(14.3\) & structures\\
\hline  
\hline 
\end{tabular}
\label{tab:significance} 
\end{center}
\end{table}
\section{Systematic uncertainties}
\label{sec:sys} 
  The systematic uncertainties considered in our analysis are the following:
  \begin{enumerate}[(i)]
    \item The uncertainty due to the CDF tracker momentum scale.
    \item The uncertainty due to the resolution model (see
          Sec.~\ref{sec:fitter}) described by the sum of two
          Gaussians. This source is expected to dominate the systematic
          uncertainties on width measurements.
    \item The choice of background model.
    \item An uncertainty due to the choice of \(Q\)-value fit range.
%
  \end{enumerate}
\par   
  To calibrate the tracker momentum scale, the energy loss in the
  material of CDF tracking detectors and the strength of the magnetic
  field must be determined.  Both effects are calibrated and analyzed in
  detail using high statistics samples of \( \jpsi \), \( \psitwos \),
  \(\OneS\), \(\Z\) reconstructed in their \(\mumu \) decay modes as
  well as \(\Dz\to\Km\pip \),
  \(\psitwos\to\jpsi(\to\mumu)\pipi\)~\cite{Acosta:2005mq,Aaltonen:2009vj}.
  The corresponding corrections are taken into account by tracking
  algorithms.
  Any systematic uncertainties on these corrections 
%
%
  are largely negligible in the \Sgbst \(Q\)-value measurements.
  The uncertainties on the measured mass differences due to the momentum
  scale
%
%
  are estimated from the deviations between \(Q_{0}\) values observed in
  similar decays reconstructed in CDF data and the known \(Q_{0}\)
  values~\cite{Nakamura:2010zzi}.  The reference modes are
  \(\Sigcpp\to\Lc\pip_{\mathit{s}}\), \(\Sigcz\to\Lc\pim_{\mathit{s}}\),
  \(\Lcst\to\Lc\pip_{\mathit{s}}\pim_{\mathit{s}}\), and
  \(\Dstarp\to\Dz\pip_{\mathit{s}}\).  
  The linear extrapolation of the measured offsets as a function of
  \(Q_{0}\) towards the \(\Sgbst\) kinematic regime is taken as the
  mass-scale uncertainty. The determined systematic uncertainty on the
  momentum scale covers also any residual charge-dependence of the
  scale.  For the mass difference \(Q_{0}\), the systematic uncertainty
  due to a possible imperfect alignment of the detector is
  negligible~\cite{Acosta:2005mq}.
\par
  Following the method used in Ref.~\cite{Abulencia:2005ry}, the
  \(\Dstarp\to\Dz(\to\Km\pip)\pip_{\mathit{s}}\) signal
  peak in the mass difference distribution \(m(\Dstarp)-m(\Dz)\) has
  been reconstructed in several bins of soft pion transverse momentum
  \(\pt(\pi_{\mathit{s}})\) starting with \(200\mevc\) as in the data.
  Each signal distribution is subjected to an unbinned maximum-likelihood fit
  with the sum of a Breit-Wigner function convoluted with a double
  Gaussian function to describe the detector resolution. The background
  under the \Dstarp signals is described by an empirical
  function~\cite{Brun:1997pa,Antcheva:2009zz}.  
  For each of the \(\pt(\pi_{\mathit{s}})\) bins, the fit determines
  the \(\Dstarp\) width, which never exceeds \( 0.2\mevcc \). Because the
  \(\Dstarp \) natural width is much smaller than the tracking
  resolution, the value of \( 0.2\mevcc \) is assigned as a systematic
  uncertainty on the measured \Sgbst natural width due to the momentum
  scale of the CDF tracker.
\par 
  Unless otherwise specified, the systematic uncertainties discussed
  below are evaluated for the measurable quantities \(Q_{0}\) and
  \(\Gamma_{0}\) by generation of statistical trials.  In each trial,
  the sample is generated according to the PDF (see
  Table~\ref{tab:fitresults}) with the nuisance parameters modified by
  the uncertainty with respect to the default set of parameters.  Then
  the sample is subjected to the unbinned maximum-likelihood fit twice,
  with the default PDF and with the PDF of the modified nuisance
  parameter set. The fit results are compared on a trial-by-trial
  basis, and their difference is computed.  The systematic uncertainty
  is found from the mean of a Gaussian fit of the distribution of the
  computed differences.
\par 
  The statistical uncertainties on the resolution model parameters due
  to the finite size of the Monte Carlo datasets introduce a systematic
  uncertainty.  Variations of the double Gaussian widths \(\sigma_{n} \)
  and \(\sigma_{w} \) and the weight \(g_{n}\) within their statistical
  uncertainties returned from the fits of Monte Carlo spectra are propagated
  into the measurable quantities using the statistical trials.
\par
\begin{figure*}  
\begin{center}
  \includegraphics[width=0.495\textwidth]
  {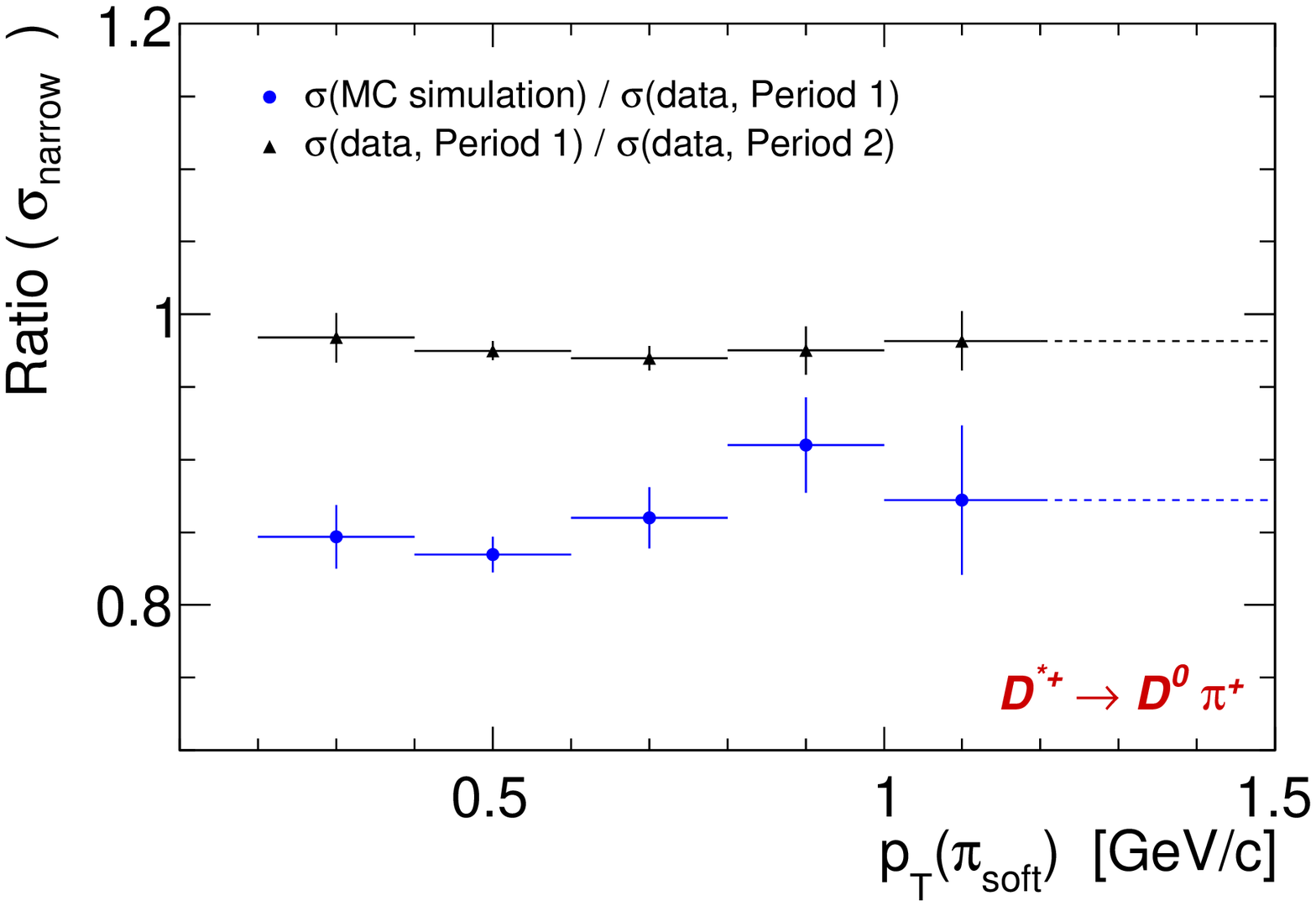}
  \includegraphics[width=0.495\textwidth]
  {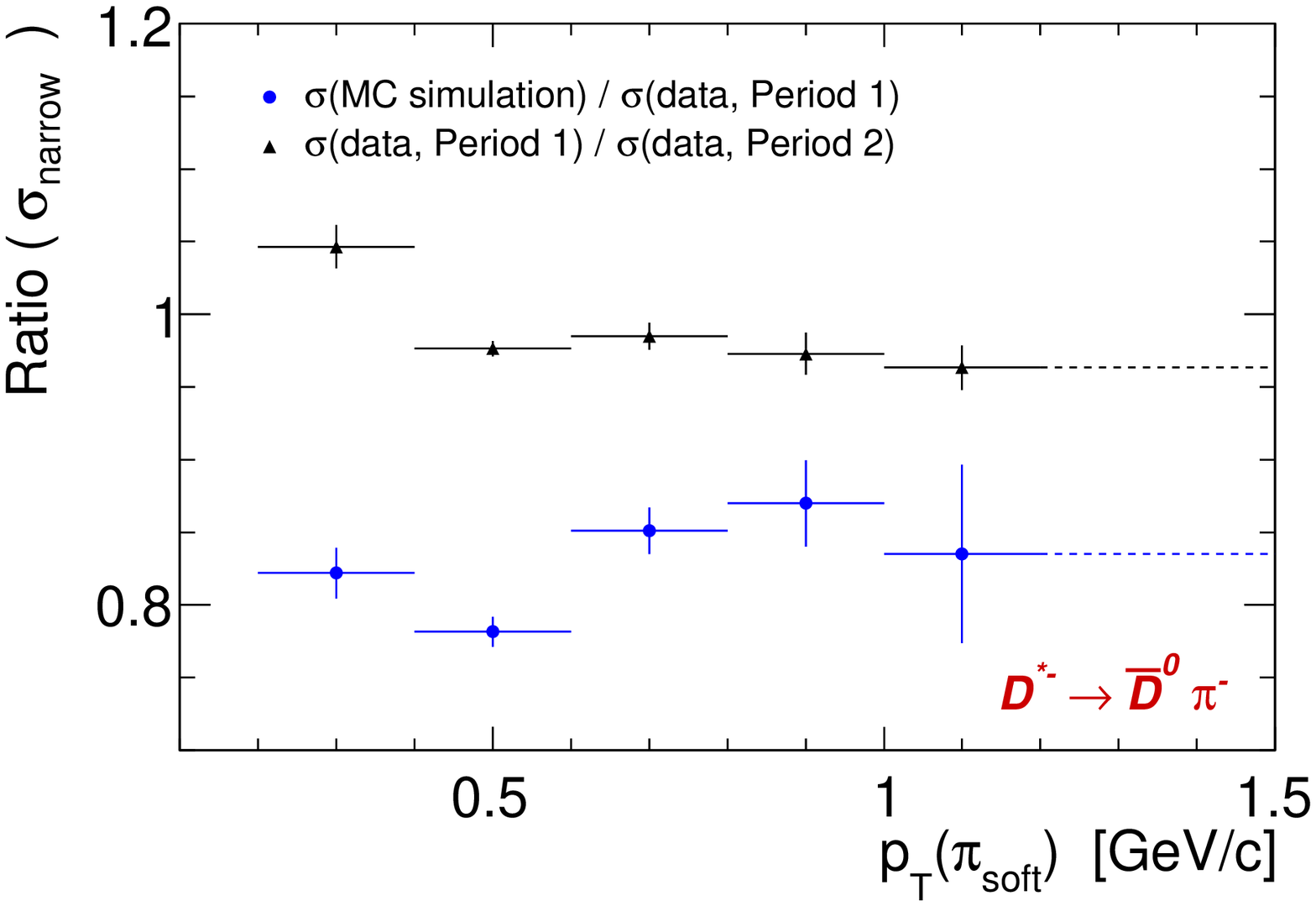}
\caption{ The left (right) plot shows the ratio of the widths of the narrow
          component of the \(\Dstarp \) (\(\Dstarm \)) mass resolution
          for data and simulation (circles) and for different
          subsamples of data (triangles) as a function of the transverse
          momentum of the soft pion. The last bin on every plot corresponds to a 
          statistics integrated above \(1.0\gevc\).  }
\label{fig:res-ratios}
\end{center}
\end{figure*}

 
   The CDF tracking simulation does not reproduce with perfect accuracy
   the tracking resolutions, especially for soft tracks at the kinematic
   threshold of \Sgbst decays.
   To estimate this contribution, we use the \(\Dstarpm\) meson decay as
   the reference mode reconstructed down to
   \(\pt(\pipm_{\mathit{s}})=200\mevc \) in the observed and simulated
   samples. We compare the mass resolution of the reference signal found
   in data with the one predicted by Monte Carlo simulation.  The
   comparison is made independently for
   \(\Dstarp\to\Dz\pip_{\mathit{s}}\) and
   \(\Dstarm\to\Dzb\pim_{\mathit{s}}\) states, as a function of soft
   pion \(\pt\) 
   using early data (Period 1) and late data (Period 2).
%
   Figure~\ref{fig:res-ratios} shows the comparisons of the narrow core
   resolution between the data and Monte Carlo both for \(\Dstarp\)
   (left plot) and \(\Dstarm\) (right plot).  The resolution
   is stable as a function of data-taking time.
\par 
  The CDF Monte Carlo simulation typically underestimates the
  \(\Dstarpm\) resolutions in the experimental data:
  \(\sigma_{n}(\mathrm{data})\,\lsim1.25\,\sigma_{n}(\mathrm{Monte~Carlo}) \). 
  Similar relations are found for the broad component of the resolution:
  \(\sigma_{w}(\mathrm{data})\,\lsim1.40\,\sigma_{w}(\mathrm{Monte~Carlo}) \). 
  These factors are used as the sources of
  the systematic uncertainties.
  The resolution extracted for the \Dstarm is systematically smaller than
  for the \Dstarp by at most \(20\% \) for \(\sigma_{n}\) and by at most
  \(40\% \) for \(\sigma_{w}\). The Monte Carlo predictions for
  \(\sigma_{n}\) and \(\sigma_{w}\) are decreased by these latter
  factors to estimate the other bounds of the systematic uncertainties. In
  both cases the conservative approach is taken.
\par 
  To find the systematic uncertainty associated with the choice of
  background shape, we change our background PDF to the one used for the
  \(\Dstarpm\) mass difference
  spectra~\cite{Brun:1997pa,Antcheva:2009zz} and compare with the
  default background PDF.
\par 
  The uncertainty associated with the fit range is estimated by varying
  the default low edge down to \(0.0015\gevcc \) and up to \(0.006\gevcc\).
  The fit results are slightly sensitive to the choice of the low
  edge and any observed biases are assigned as another systematic
  uncertainty. 
\par 
  The final systematic uncertainties are listed in
  Table~\ref{tab:finalSystematics}.
\begin{table*}
\begin{center}
\caption{ Summary of the systematic uncertainties listed in the
          following order: mass scale, resolution, choice of background
          model, and fit range. The total systematic uncertainty is
          obtained by adding all the associated uncertainties in
          quadrature. The last column shows the percentage of the total
          systematic uncertainty relative to its central value. }
\begin{tabular}{lcccccc}
\hline
\hline
 Measurable quantity & Scale & Resolution & Background & Fit range & Total & Percentage \\ 
\hline 
\multirow{2}{*}{\(Q(\Sigbm)\,[\mevcc]\)} &           & \(+0.06\)  & \(+0.04\) & \(+0.02\) 
                                                                  & \(+0.07\) & \(+0.1\) \\ 
                                         & \(-0.38\) & \(-0.07\)  & \(-0.04\) & \(-0.03\) 
                                                                  & \(-0.39\) & \(-0.7\) \\ 
\hline 
\multirow{2}{*}{\(\Gamma(\Sigbm)\,[\mevcc]\)} & \(+0.20\)  & \(+0.85\) & \(+0.50\) & \(+0.50\) 
                                                                       & \(+1.13\) & \(+23\) \\ 
                                              & \(-0.20\)  & \(-0.87\) & \(-0.50\) & \(-0.51\) 
                                                                       & \(-1.14\) & \(-23\) \\ 
\hline 
%
\multirow{2}{*}{\(Q(\Sigbstm)\,[\mevcc]\)} &           & \(+0.06\) & \(+0.06\) & \(+0.02\) 
                                                                   & \(+0.09\) & \(+0.1\) \\ 
                                           & \(-0.56\) & \(-0.08\) & \(-0.06\) & \(-0.09\) 
                                                                   & \(-0.58\) & \(-0.8\) \\ 
\hline 
\multirow{2}{*}{\(\Gamma(\Sigbstm)\,[\mevcc]\)} & \(+0.20\) & \(+0.65\) & \(+0.30\) & \(+0.50\) 
                                                                        & \(+0.89\) & \(+12\) \\ 
                                                & \(-0.20\) & \(-0.96\) & \(-0.30\) & \(-0.90\) 
                                                                        & \(-1.36\) & \(-18\) \\ 
\hline 
%
\multirow{2}{*}{\(Q(\Sigbp)\,[\mevcc]\)} &           &  \(+0.07\) & \(+0.05\) & \(+0.02\)  
                                                                  & \(+0.09\) & \(+0.2\) \\ 
                                         & \(-0.35\) &  \(-0.12\) & \(-0.05\) & \(-0.03\)  
                                                                  & \(-0.38\) & \(-0.7\) \\ 
\hline 
\multirow{2}{*}{\(\Gamma(\Sigbp)\,[\mevcc]\)} & \(+0.20\) & \(+0.94\) & \(+0.40\) & \(+0.50\) 
                                                                      & \(+1.16\) & \(+12\) \\ 
                                              & \(-0.20\) & \(-0.90\) & \(-0.40\) & \(-0.51\) 
                                                                      & \(-1.12\) & \(-12\) \\ 
\hline 
%
\multirow{2}{*}{\(Q(\Sigbstp)\,[\mevcc]\)} &           & \(+0.06\) & \(+0.10\) & \(+0.02\) 
                                                                   & \(+0.12\) & \(+0.2\) \\ 
                                           & \(-0.52\) & \(-0.13\) & \(-0.10\) & \(-0.09\) 
                                                                   & \(-0.55\) & \(-0.8\) \\ 
\hline 
\multirow{2}{*}{\(\Gamma(\Sigbstp)\,[\mevcc]\)} & \(+0.20\)  &  \(+0.64\) & \(+0.50\) & \(+0.50\) 
                                                                          & \(+0.97\) & \(+8.5\) \\ 
                                                & \(-0.20 \) &  \(-1.01\) & \(-0.50\) & \(-0.90\) 
                                                                          & \(-1.46\) & \(-13\) \\ 
\hline 
\hline 
\end{tabular}
\label{tab:finalSystematics}
\end{center}
\end{table*}
\section{Results and conclusions}
\label{sec:results}
  The analysis results are arranged in Table~\ref{tab:results}.
  From the measured \Sgbstpm \(Q\) values we extract the absolute masses
  using the known value of the \(\pi^{\pm}\)
  mass~\cite{Nakamura:2010zzi} and the CDF  \Lb  mass measurement,
  \(m(\Lb) = 5619.7\pm1.2\,\stat\pm1.2\,\syst\,\mevcc\,\),
  as obtained in an independent sample~\cite{Acosta:2005mq}.
%
%
  The \Lb statistical and systematic uncertainties contribute to the
  systematic uncertainty  on the \Sgbstpm absolute masses.
\par 
  Using the measured \(Q\) values, we extract the isospin mass splittings
  for the isotriplets of the \( J^{P}={\frac{1}{2}}^{+} \) and
  \(J^{P}={\frac{3}{2}}^{+}\) states. 
  The statistical uncertainties on the \(Q\)-measurements of the
  corresponding charge states are added in quadrature.  We assume that
  the correlated systematic uncertainties due to mass scale, fit bias
  due to choice of fit range, and imperfect Monte Carlo description of
  the resolution are completely canceled in the isospin mass splittings.
  The uncertainties due to background choice are added in quadrature.
\begin{table*}
\begin{center}
\caption{ Summary of the final results.  The first uncertainty is 
          statistical and the second is systematic. } 
\begin{tabular}{lccc}
\hline
\hline
State        & \(Q\) value,\mevcc & Absolute mass \(m\),\mevcc  & Natural width \(\Gamma\),\mevcc \\
\hline
\multirow{2}{*}{\Sigbm} & \multirow{2}{*}{\(56.2\,_{-0.5\,-0.4}^{+0.6\,+0.1} \)} 
                        & \multirow{2}{*}{\(5815.5\,_{-0.5}^{+0.6}\pm1.7 \)} 
                        & \multirow{2}{*}{\(4.9\,_{-2.1}^{+3.1}\pm{1.1} \)} \\
                        &                                    
                        &                                   
                        & \\
\multirow{2}{*}{\Sigbstm} & \multirow{2}{*}{\(75.8\,\pm0.6\,_{-0.6}^{+0.1} \)} 
                          & \multirow{2}{*}{\(5835.1\,\pm0.6\,_{-1.8}^{+1.7} \)} 
                          & \multirow{2}{*}{\(7.5\,_{-1.8\,-1.4}^{+2.2\,+0.9} \)} \\
                          &                                    
                          &                                   
                          & \\
\multirow{2}{*}{\Sigbp} & \multirow{2}{*}{\(52.1\,_{-0.8\,-0.4}^{+0.9\,+0.1} \)} 
                        & \multirow{2}{*}{\(5811.3\,_{-0.8}^{+0.9}\pm1.7 \)} 
                        & \multirow{2}{*}{\(9.7\,_{-2.8\,-1.1}^{+3.8\,+1.2} \)} \\
                        &             
                        &                                 
                        & \\
\multirow{2}{*}{\Sigbstp} & \multirow{2}{*}{\(72.8\,\pm0.7\,_{-0.6}^{+0.1} \)} 
                          & \multirow{2}{*}{\(5832.1\,\pm0.7\,_{-1.8}^{+1.7}  \)} 
                          & \multirow{2}{*}{\(11.5\,_{-2.2\,-1.5}^{+2.7\,+1.0} \)} \\
                          &                                    
                          &                                   
                          & \\
\hline
%
  & \multicolumn{3}{c}{Isospin mass splitting, \mevcc} \\
\hline
%
 \multirow{2}{*}{\(m(\Sigbp) - m(\Sigbm)\)} 
  & \multicolumn{3}{c}{ \multirow{2}{*}{\( -4.2\,_{-1.0}^{+1.1}\pm{0.1} \)} } \\
                          &                                    
                          &                                   
                          & \\
 \multirow{2}{*}{\(m(\Sigbstp) - m(\Sigbstm)\)} 
  & \multicolumn{3}{c}{ \multirow{2}{*}{\( -3.0\,^{+1.0}_{-0.9}\pm{0.1} \)} } \\
                          &                                    
                          &                                   
                          & \\
\hline
\hline
\end{tabular}
\label{tab:results}
\end{center}
\end{table*}
%
%
%
\par
  In conclusion, we have measured the masses and widths of the
  \(\Sgbstpm\) baryons using a sample of approximately \(16~300\)
  \(\Lb\) candidates reconstructed in their \(\Lb\to\Lc\pim\) mode
  corresponding to \( 6\invfb \) of CDF data.
\par
  The first observation~\cite{:2007rw} of the \(\Sgbstpm\) bottom
  baryons has been confirmed with every individual signal reconstructed
  with a significance well in excess of six Gaussian standard
  deviations.
\par
  The statistical precision on the direct mass differences is improved
  by a factor of two over the previous measurement~\cite{:2007rw}.  The
  measurements are in good agreement with the previous results and
  supersede them.
\par
  The isospin mass splittings within the \( I=1 \) triplets of the \( \Sigb \)
  and \( \Sigbst \) states have been extracted for the first time.
  The \(\Sgbstm\) states have higher masses than their \(\Sgbstp\)
  partners, following a pattern common to most of the
  known isospin multiplets~\cite{Chan:1985ty}. This  measurement favors the
  phenomenological explanation of this ordering as due to the higher masses of
  the \d quark with respect to the \u quark and the larger electromagnetic
  contribution due to
  electrostatic Coulomb forces between quarks in \(\Sgbstm\) states than in
  \(\Sgbstp\) ones.
  The difference in the measured isospin mass splittings between 
  the \(\Sigbst\) and \(\Sigb\) isotriplets supports the theoretical
  estimate of Ref.~\cite{Rosner:2006yk}.
  The natural widths of the \( \Sigbpm \) and \( \Sigbstpm \) states have 
  been measured for the first time. The measurements are in agreement 
  with theoretical expectations.
\begin{acknowledgments}
  We thank the Fermilab staff and the technical staffs of the
  participating institutions for their vital contributions. This work
  was supported by the U.S. Department of Energy and National Science
  Foundation; the Italian Istituto Nazionale di Fisica Nucleare; the
  Ministry of Education, Culture, Sports, Science and Technology of
  Japan; the Natural Sciences and Engineering Research Council of
  Canada; the National Science Council of the Republic of China; the
  Swiss National Science Foundation; the A.P. Sloan Foundation; the
  Bundesministerium f\"ur Bildung und Forschung, Germany; the Korean
  World Class University Program, the National Research Foundation of
  Korea; the Science and Technology Facilities Council and the Royal
  Society, UK; the Russian Foundation for Basic Research; the Ministerio
  de Ciencia e Innovaci\'{o}n, and Programa Consolider-Ingenio 2010,
  Spain; the Slovak R\&D Agency; the Academy of Finland; and the
  Australian Research Council (ARC). 
\end{acknowledgments}

\begin{thebibliography}{99}
%
%
\bibitem{Isgur:1989vq}
  N.~Isgur and M.~B.~Wise,
  Phys.\ Lett.\  B {\bf 232}, 113 (1989);
%
  {\it Ibid.} {\bf 237}, 527 (1990);
%
  N.~Isgur and M.~B.~Wise,
  Phys.\ Rev.\  D {\bf 42}, 2388 (1990).
%
\bibitem{Neubert:1993mb}
  M.~Neubert,
  Phys.\ Rept.\  {\bf 245}, 259 (1994).
%
\bibitem{Manohar:2000dt}
  A.~V.~Manohar and M.~B.~Wise,
  Camb.\ Monogr.\ Part.\ Phys.\ Nucl.\ Phys.\ Cosmol.\  {\bf 10}, 1 (2000).
%
\bibitem{Korner:1994nh}
  J.~G.~Korner, M.~Kramer, and D.~Pirjol,
  Prog.\ Part.\ Nucl.\ Phys.\  {\bf 33}, 787 (1994).
%
\bibitem{notation:sgbst}
  Throughout the text the notations \Sgbst or \Sgbstpm
  represent the four states \Sigbm, \Sigbstm, \Sigbp and \Sigbstp while
  the notations \Sgbstm or \Sgbstp include \Sigbm, \Sigbstm or \Sigbp,
  \Sigbstp pairs of states correspondingly.
%
\bibitem{Nakamura:2010zzi}
  K.~Nakamura {\it et al.}  (Particle Data Group),
  J.\ Phys.\ G {\bf 37}, 075021 (2010).
%
\bibitem{Ebert:2005xj}
D.~Ebert, R.~N.~Faustov, and V.~O.~Galkin,
 Phys.\ Rev.\  D {\bf 72}, 034026 (2005);
%
  D.~Ebert, R.~N.~Faustov, and V.~O.~Galkin,
  Phys.\ Lett.\  B {\bf 659}, 612 (2008);
%
  D.~Ebert, R.~N.~Faustov, and V.~O.~Galkin,
  Phys.\ Atom.\ Nucl.\  {\bf 72}, 178 (2009).
%
\bibitem{Jenkins:1996de}
  E.~E.~Jenkins,
  Phys.\ Rev.\  D {\bf 54}, 4515 (1996);
%
  {\it Ibid.} {\bf 55}, 10 (1997);
%
  {\it Ibid.} {\bf 77}, 034012 (2008).
%
\bibitem{Rosner:2006jz}
  J.~L.~Rosner,
  J.\ Phys.\ G {\bf 34}, S127 (2007);
%
  M.~Karliner and H.~J.~Lipkin,
  Phys.\ Lett.\  B {\bf 660}, 539 (2008);
%
%
  M.~Karliner, B.~Keren-Zur, H.~J.~Lipkin, and J.~L.~Rosner,
  Annals Phys.\  {\bf 324}, 2 (2009);
%
  M.~Karliner,
  Nucl.\ Phys.\ Proc.\ Suppl.\  {\bf 187}, 21 (2009);
%
  H.~Garcilazo, J.~Vijande, and A.~Valcarce,
  J.\ Phys.\ G {\bf 34}, 961 (2007).
%
\bibitem{Rosner:2006yk}  
J.~L.~Rosner,  
Phys.\ Rev.\  D {\bf 75}, 013009 (2007).  
%
\bibitem{Roncaglia:1994ex}
  R.~Roncaglia, A.~Dzierba, D.~B.~Lichtenberg, and E.~Predazzi,
  Phys.\ Rev.\  D {\bf 51}, 1248 (1995);
%
  R.~Roncaglia, D.~B.~Lichtenberg, and E.~Predazzi,
  Phys.\ Rev.\  D {\bf 52}, 1722 (1995);
%
  D.~B.~Lichtenberg, R.~Roncaglia, and E.~Predazzi,
  Phys.\ Rev.\  D {\bf 53}, 6678 (1996).
%
\bibitem{Liu:2007fg}
  X.~Liu, H.~X.~Chen, Y.~R.~Liu, A.~Hosaka, and S.~L.~Zhu,
  Phys.\ Rev.\  D {\bf 77}, 014031 (2008);
%
  J.~R.~Zhang and M.~Q.~Huang,
  Phys.\ Rev.\  D {\bf 77}, 094002 (2008).
%
\bibitem{Mathur:2002ce}
  N.~Mathur, R.~Lewis, and R.~M.~Woloshyn,
  Phys.\ Rev.\  D {\bf 66}, 014502 (2002);
%
  R.~Lewis and R.~M.~Woloshyn,
  Phys.\ Rev.\  D {\bf 79}, 014502 (2009).
%
\bibitem{Isgur:1979ed} 
  N.~Isgur,
  Phys.\ Rev.\ D {\bf 21}, 779 (1980)
  [Erratum-ibid.\ D {\bf 23}, 817 (1981)];
%
  N.~Isgur, H.~R.~Rubinstein, A.~Schwimmer and H.~J.~Lipkin,
  Phys.\ Lett.\ B {\bf 89}, 79 (1979);
%
  S.~Godfrey and N.~Isgur,
  Phys.\ Rev.\ D {\bf 34}, 899 (1986).
%
%
\bibitem{Chan:1985ty}
  L.~H.~Chan,
  Phys.\ Rev.\  D {\bf 31}, 204 (1985);
%
  W.~Y.~P.~Hwang and D.~B.~Lichtenberg,
  Phys.\ Rev.\  D {\bf 35}, 3526 (1987);
%
  S.~Capstick,
  Phys.\ Rev.\  D {\bf 36}, 2800 (1987);
%
  M.~Genovese, J.~M.~Richard, B.~Silvestre-Brac, and K.~Varga, 
  Phys.\ Rev.\  D {\bf 59}, 014012 (1998);
%
  C.~W.~Hwang and C.~H.~Chung,
  Phys.\ Rev.\  D {\bf 78}, 073013 (2008).
%
%
\bibitem{Capstick:2000qj}
  S.~Capstick and W.~Roberts,
  Prog.\ Part.\ Nucl.\ Phys.\  {\bf 45}, S241 (2000).
%
\bibitem{Guo:2007qu}
  X.~H.~Guo, K.~W.~Wei, and X.~H.~Wu,
  Phys.\ Rev.\  D {\bf 77}, 036003 (2008).
%
\bibitem{Hwang:2006df}
  C.~W.~Hwang,
  Eur.\ Phys.\ J.\  C {\bf 50}, 793 (2007).
%
\bibitem{:2007rw}
  T.~Aaltonen {\it et al.} (CDF Collaboration),
  Phys.\ Rev.\ Lett.\  {\bf 99}, 202001 (2007).
%
\bibitem{:2007ub}
   V.~M.~Abazov {\it et al.} (\dzero Collaboration),
   Phys.\ Rev.\ Lett.\  {\bf 99}, 052001 (2007).
%
 \bibitem{:2007un}
   T.~Aaltonen {\it et al.} (CDF Collaboration),
   Phys.\ Rev.\ Lett.\  {\bf 99}, 052002 (2007).
 %
 \bibitem{Aaltonen:2009ny}
   T.~Aaltonen {\it et al.} (CDF Collaboration),
   Phys.\ Rev.\ D {\bf 80}, 072003 (2009).
 %
 \bibitem{Abazov:2008qm}
   V.~M.~Abazov {\it et al.} (\dzero Collaboration),
   Phys.\ Rev.\ Lett.\  {\bf 101}, 232002 (2008).
 %
\bibitem{Aaltonen:2011wd}
  T.~Aaltonen {\it et al.} (CDF Collaboration),
  Phys.\ Rev.\ Lett.\  {\bf 107}, 102001 (2011).
%
\bibitem{Aaltonen:2011sf}
  T.~Aaltonen {\it et al.} (CDF Collaboration),
  Phys.\ Rev.\ D {\bf 84}, 012003 (2011).
%
%
%
\bibitem{Acosta:2004yw}
  D.~Acosta {\it et al.}  (CDF Collaboration),
  Phys.\ Rev.\  D {\bf 71}, 032001 (2005).
%
\bibitem{Sill:2000zz}
A.~Sill {\it et al.},
Nucl.\ Instrum.\ Meth.\ A {\bf 447}, 1 (2000).
%
\bibitem{Affolder:2000tj}
  A.~A.~Affolder {\it et al.} (CDF Collaboration),
  Nucl.\ Instrum.\ Meth.\  A {\bf 453}, 84-88 (2000).
%
\bibitem{Nahn:2003tm}
  S.~Nahn  (On behalf of the CDF Collaboration),
  Nucl.\ Instrum.\ Methods\  A {\bf 511}, 20 (2003).
%
\bibitem{Hill:2004qb}
  C.~S.~Hill  (On behalf of the CDF Collaboration),
  Nucl.\ Instrum.\ Methods\  A {\bf 530}, 1 (2004).
%
\bibitem{Affolder:2003ep}
  A.~A.~Affolder {\it et al.}  (CDF Collaboration),
  Nucl.\ Instrum.\ Methods\  A {\bf 526}, 249 (2004).
%
\bibitem{Thomson:2002xp}
  E.~J.~Thomson {\it et al.},
  IEEE Trans.\ Nucl.\ Sci.\  {\bf 49}, 1063 (2002).
%
\bibitem{Ashmanskas:2003gf}
  B.~Ashmanskas {\it et al.} (CDF Collaboration),
  Nucl.\ Instrum.\ Meth.\ A {\bf 518}, 532-536 (2004).
%
\bibitem{RistoriPunzi:CDFTrigger}
  L.~Ristori and G.~Punzi,
   Ann.\ Rev.\ Nucl.\ Part.\ Sci.\ {\bf 60}, 595-614 (2010).
%
%
\bibitem{Nason:1987xz}
  P.~Nason, S.~Dawson, and R.~K.~Ellis,
  Nucl.\ Phys.\ B {\bf 303}, 607 (1988);
%
  {\it Ibid.},
  Nucl.\ Phys.\ B {\bf 327}, 49-92 (1989).
%
\bibitem{Peterson:1982ak}
  C.~Peterson, D.~Schlatter, I.~Schmitt, and P.~M.~Zerwas,
  Phys.\ Rev.\ D {\bf 27}, 105 (1983). 
%
\bibitem{Lange:2001uf}
  D.~J.~Lange,
  Nucl.\ Instrum.\ Methods\  A {\bf 462}, 152 (2001).
%
\bibitem{Geant} 
  R. Brun, R. Hagelberg, M. Hansroul, and J.C. Lassalle, CERN Reports
  No. CERN-DD-78-2-REV and No. CERN-DD-78-2.
%
\bibitem{notation:cc}
  Unless otherwise stated all references to a specific charge
  combination imply the charge conjugate combination as well.
  Specifically, \(\SgbstmBar\to\LbBar\pim_{\mathit{s}}\),
  \(\SgbstpBar\to\LbBar\pip_{\mathit{s}}\),
  \( \LbBar\to\Lcbar\pip_{b} \), \(\Lcbar\to\overline{\proton}\Kp\pim \),
  \(\Dstarm\to\Dzb(\to\Kp\pim)\pim_{\mathit{s}}\).
%
\bibitem{Abulencia:2006df}
  A.~Abulencia {\it et al.}  (CDF Collaboration),
  Phys.\ Rev.\ Lett.\  {\bf 98}, 122002 (2007).
%
\bibitem{Jackson:1964zd}
    J.~D.~Jackson, Nuovo Cim. {\bf 34}, 1644 (1964).
%
%
 \bibitem{Wilks} S.S.~Wilks, 
 Ann.\ Math.\ Statist.\ {\bf 9} (1938) 60-2.
%
\bibitem{Royall}
	R.~Royall,
	J.\ Amer.\ Statist.\ Assoc.\ {\bf 95}, 760 (2000).
%
%
%
\bibitem{Acosta:2005mq} 
  D.~Acosta {\it et al.}  (CDF Collaboration),
  Phys.\ Rev.\ Lett.\  {\bf 96}, 202001 (2006).
%
\bibitem{Aaltonen:2009vj} 
  T.~Aaltonen {\it et al.}  (CDF Collaboration),
  Phys.\ Rev.\ Lett.\  {\bf 103}, 152001 (2009)
%
\bibitem{Abulencia:2005ry}
  A.~Abulencia {\it et al.}  (CDF Collaboration),
  Phys.\ Rev.\  D {\bf 73}, 051104 (2006).
%
\bibitem{Brun:1997pa}
  R.~Brun and F.~Rademakers,
  Nucl.\ Instrum.\ Methods\  A {\bf 389}, 81 (1997);
  see also {\tt http://root.cern.ch/}.
%
\bibitem{Antcheva:2009zz}
  I.~Antcheva {\it et al.},
  Comput.\ Phys.\ Commun.\  {\bf 180}, 2499 (2009).
%
\bibitem{CalanchaParedes:2011zz}
  C.~Calancha Paredes, 
  FERMILAB-THESIS-2011-19, Feb 2011, 218 pp.
%
\end{thebibliography}
\end{document}